%% file: ms.tex
\documentclass[10pt,journal,compsoc]{IEEEtran}
\newif\ifpeerreview

\peerreviewfalse

\usepackage[nocompress]{cite}
\usepackage{url}
\usepackage{amsmath,amssymb,graphicx}
\usepackage{multirow}
\usepackage[dvipsnames]{xcolor}
\usepackage{lipsum} % Only used to generate random text.
\usepackage{hyperref}
\usepackage{tcolorbox}
\usepackage[figuresright]{rotating}
\usepackage[switch]{lineno}

\newcommand{\fangzhou}[1]{\textcolor{black}{~{#1}}}

\newcommand{\paperID}{61}

\title{Physics to the Rescue: Deep Non-line-of-sight Reconstruction for High-speed Imaging}

\author{Fangzhou~Mu,
        Sicheng~Mo,
        Jiayong~Peng,
        Xiaochun~Liu,
        Ji~Hyun~Nam,
        Siddeshwar~Raghavan,
        Andreas~Velten 
        and~Yin~Li%,~\IEEEmembership{Member,~IEEE}% <-this % stops a space
\IEEEcompsocitemizethanks{\IEEEcompsocthanksitem F. Mu, S. Mo, X. Liu, J. H. Nam, S. Raghavan, A. Velten and Y. Li are with the University of Wisconsin-Madison, Madison,
WI, 53706, United States. E-mail: \{fmu2, smo3, xliu669, jnam26, sraghavan7, velten, yin.li\}@wisc.edu.}
\IEEEcompsocitemizethanks{\IEEEcompsocthanksitem J. Peng is with the University of Science and Technology of China, Hefei, Anhui Province, 230026, China. E-mail: jiayong@mail.ustc.edu.cn.}%
}
\begin{document}

\IEEEtitleabstractindextext{%
\begin{abstract}
% We address the challenging problem of non-line-of-sight reconstruction using a non-confocal rapid imaging acquisition system. The key challenge lies in the approximation of light paths during non-confocal imaging, necessary for the fast capturing. To this end, we present a novel deep model that incorporates physics models of inverse feature propagation and volume rendering into a neural network for reconstruction. Our orchestrated design leads to a method that can be trained using various supervision signals including intensity images or transient measurements, is able to render both intensity and depth images at inference time with a single forward pass, and achieves superior results for both synthetic data and real captures.
\fangzhou{Computational approach to imaging around the corner, or non-line-of-sight (NLOS) imaging, is becoming a reality thanks to major advances in imaging hardware and reconstruction algorithms. A recent development towards practical NLOS imaging, Nam et al.~\cite{nam2021low} demonstrated a high-speed non-confocal imaging system that operates at 5Hz, 100x faster than the prior art. This enormous gain in acquisition rate, however, necessitates numerous approximations in light transport, breaking many existing NLOS reconstruction methods that assume an idealized image formation model.
%breaking many existing learning based NLOS reconstruction methods whose training relies on synthetic data that abide an idealized image formation model. 
To bridge the gap, we present a novel deep model that incorporates the complementary physics priors of wave propagation and volume rendering into a neural network for high-quality and robust NLOS reconstruction. This orchestrated design regularizes the solution space by relaxing the image formation model, resulting in a deep model that generalizes well on real captures despite being exclusively trained on synthetic data. Further, we devise a unified learning framework that enables our model to be flexibly trained using diverse supervision signals, including target intensity images or even raw NLOS transient measurements. Once trained, our model renders both intensity and depth images at inference time in a single forward pass, capable of processing more than 5 captures per second on a high-end GPU. Through extensive qualitative and quantitative experiments, we show that our method outperforms prior physics and learning based approaches on both synthetic and real measurements. We anticipate that our method along with the fast capturing system will accelerate future development of NLOS imaging for real world applications that require high-speed imaging.}
\end{abstract}

}

% Make Title
\ifpeerreview
\linenumbers \linenumbersep 15pt\relax 
\author{Paper ID \paperID\IEEEcompsocitemizethanks{\IEEEcompsocthanksitem This paper is under review for ICCP 2022 and the PAMI special issue on computational photography. Do not distribute.}}
\markboth{Anonymous ICCP 2022 submission ID \paperID}%
{}
\fi
\maketitle
\thispagestyle{empty}

% intro
\input{tex/01_intro}

% related works
\input{tex/02_related_works}

% problem statement
\input{tex/03_problem_statement}

% method
\input{tex/04_method}

% training & inference
\input{tex/05_training_inference}

% experiments and results
\input{tex/06_exp_results}

\section{Conclusion}
In this paper, we presented a novel learning based method for non-confocal NLOS reconstruction. Our key innovation is a method that embeds strong domain knowledge in the deep model in the form of an inverse propagation module and a volume renderer, both physics based, to navigate the learning of a conditional neural scene representation. Moreover, our model can be flexibly trained using diverse supervision signals including multi-view target images, and more importantly transient measurements themselves. We demonstrated superior reconstruction quality of our model in comparison to state-of-the-art physics and learning based methods. In particular, our method, despite being trained on synthetic data, generalizes well on real measurements. We anticipate that our method, along with the fast imaging system, will lay the foundation for exciting applications of NLOS imaging that require high-speed imaging. We hope our method will provide a solid step towards the challenging problem of NLOS reconstruction, and shed light on a broader spectrum of inverse problems in imaging sciences.

\begin{figure}[!t]
\centering
\includegraphics[width=0.9\linewidth]{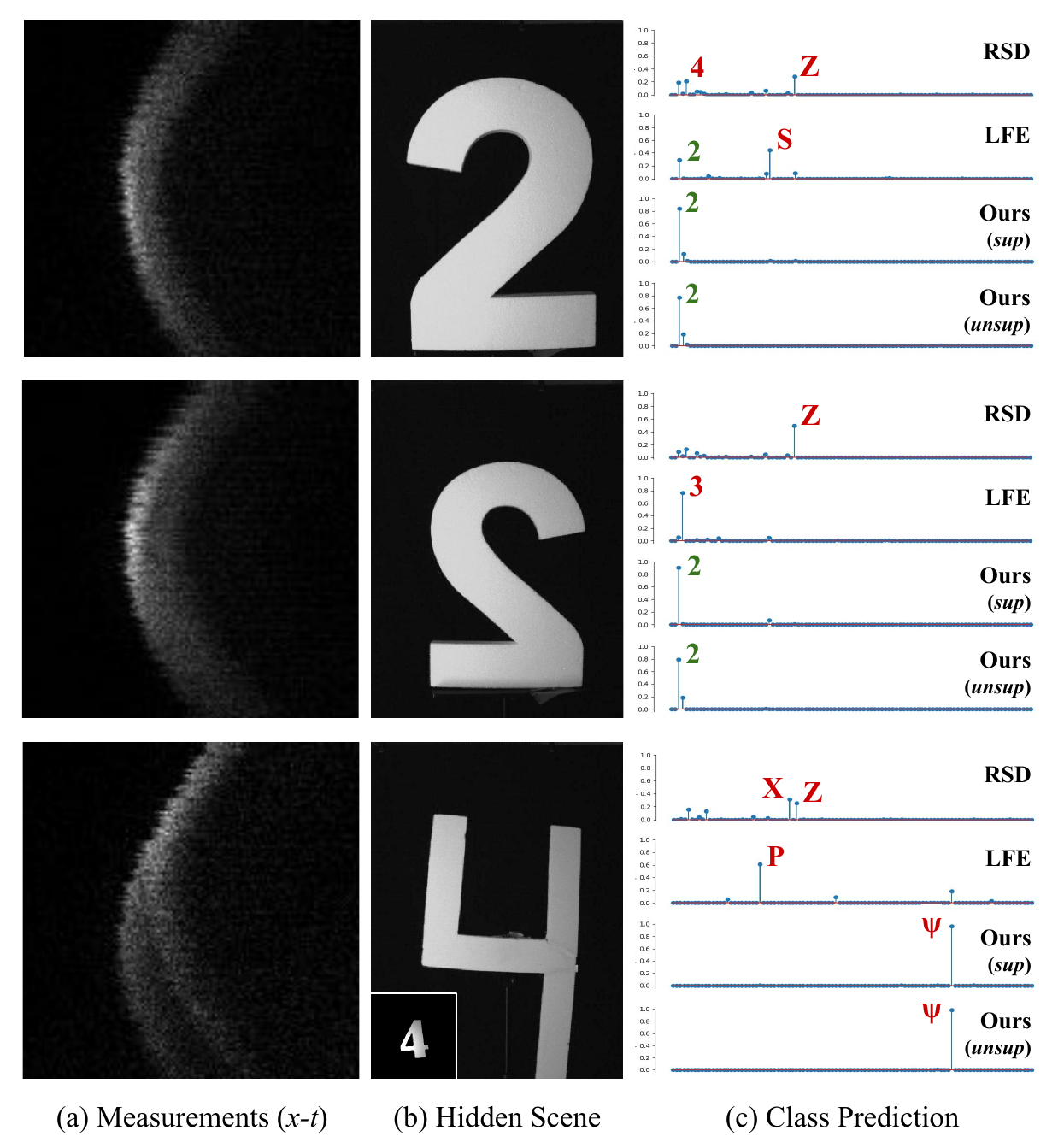} \vspace{-1em}
\caption{\textbf{Classification on real measurements.} \textbf{(a)} An \textit{x-t} slice of the input measurement volume. \textbf{(b)} A reference image of the hidden scene (not used for inference). \textbf{(c)} Predicted class probabilities. Taller lines indicate higher probability values. {\color{Green}{\textbf{Green}}} for correct predictions and {\color{Red}{\textbf{red}}} for incorrect predictions.}
\vspace{-0.5em}
\label{classification_real}
\end{figure}

\begin{figure}[!t]
\centering
\includegraphics[width=\linewidth]{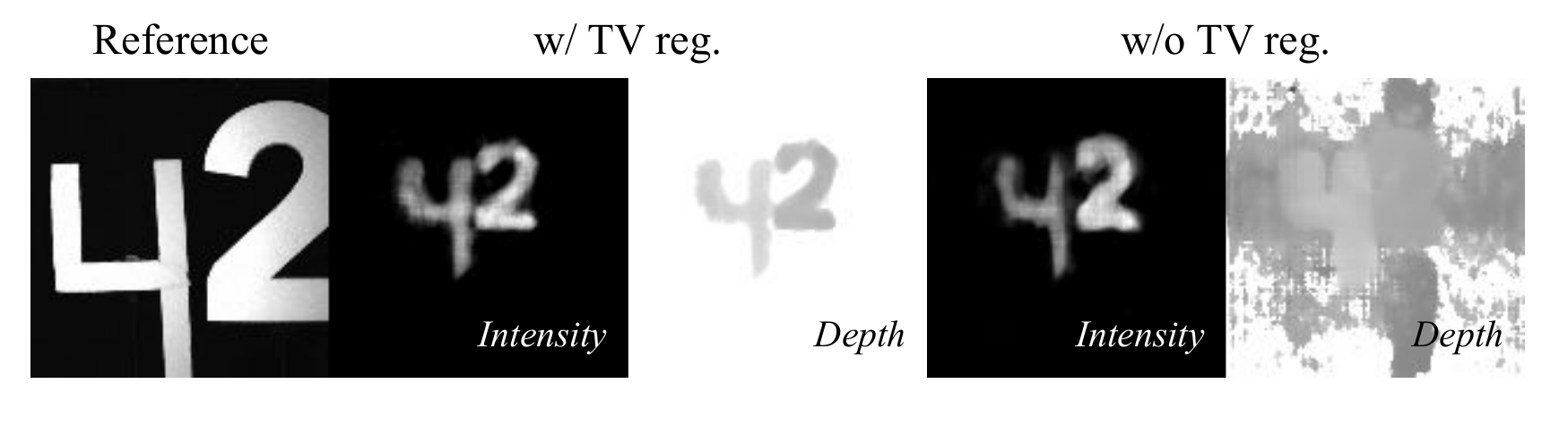} \vspace{-3em}
\caption{\fangzhou{\textbf{Ablation on total variation prior.} Our total variation regularizer eliminates spurious density in the reconstruction volume.}}
\vspace{-0.5em}
\label{tv_ablation}
\end{figure}

\begin{figure}[!t]
\centering
\includegraphics[width=\linewidth]{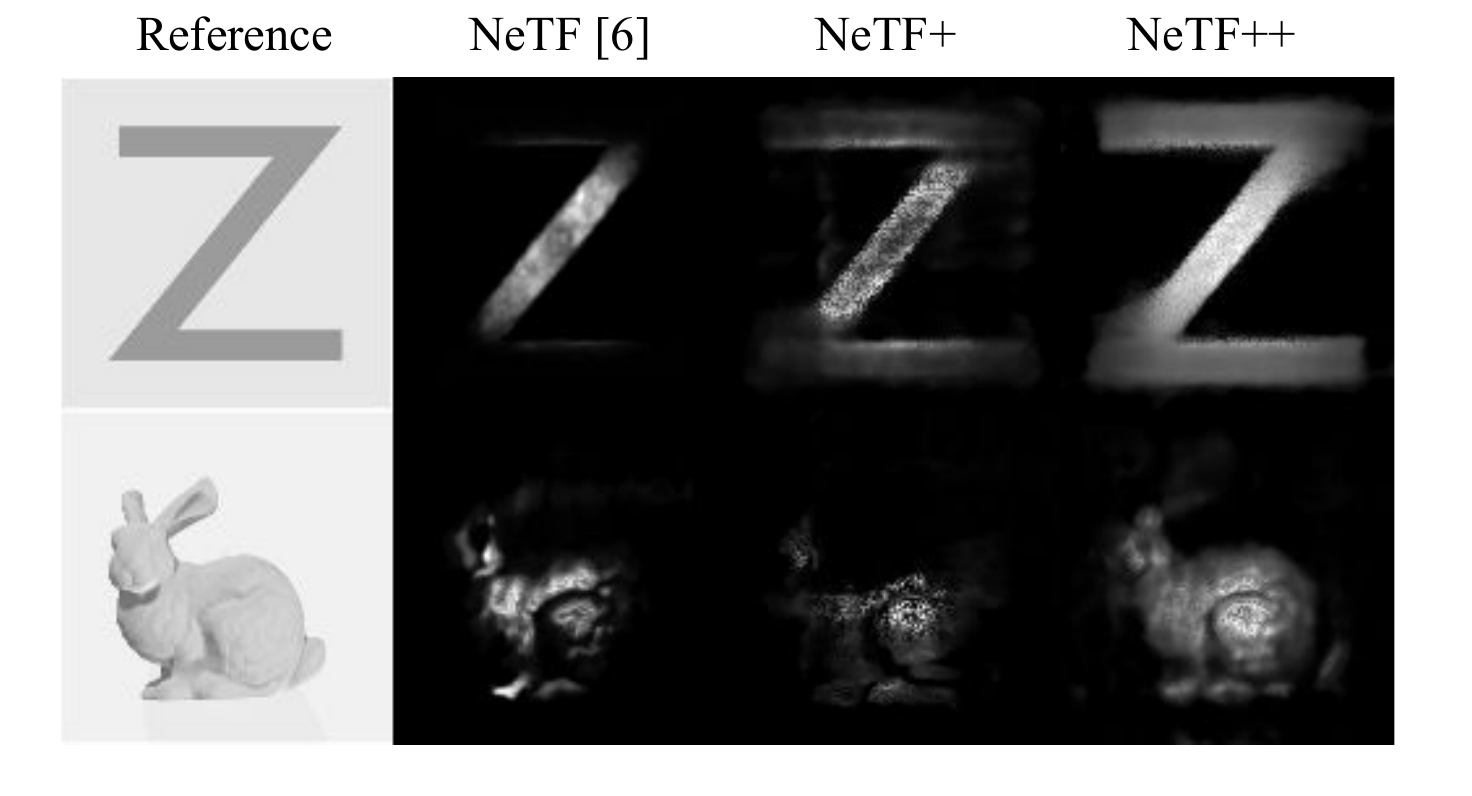} \vspace{-3em}
\caption{\fangzhou{\textbf{Ablation on transient rendering recipe.} Our formulation (NeTF++) employs a principled sampling strategy for the estimation of transmittance in the rendering equation (Figure~\ref{sampling}), whereas NeTF and NeTF+ omit transmittance and yield worse reconstruction. Note that we display raw renderings without contrast adjustment for fair comparison.}}
\vspace{-0.5em}
\label{netf_ablation}
\end{figure}

% Any acknowledgments to only be included in camera ready
\ifpeerreview \else
\section*{Acknowledgments}
The authors acknowledge support from UW VCRGE with funding from WARF, DARPA (HR0011-16-C-0025) and the Air Force Office for Scientific Research (FA9550-21-1-0341).
\fi

\bibliographystyle{IEEEtran}
\bibliography{references}

\ifpeerreview \else
%%%% For the camera ready version, please fill out this
%%%% biography. Your camera ready should be within a 12 page limit
%%%% including acknowledgments, references and biography.

% \includegraphics command within an optional argument. (You could
% create your own custom macro containing the \includegraphics command
% to make things simpler here.)

\vspace{-2em}
\begin{IEEEbiography}[{\includegraphics[width=1in,height=1.25in,clip,keepaspectratio]{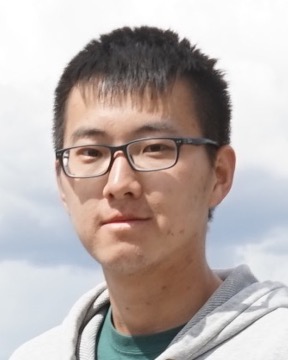}}]{Fangzhou Mu} is a fourth-year PhD student in Computer Sciences at the University of Wisconsin-Madison, advised by Professor Yin Li. His research interest lies in computer vision, computational photography and deep learning. He received his Master's degree from UW-Madison and his Bachelor's degree from Zhejiang University.
\end{IEEEbiography}\vspace{-3em}

\begin{IEEEbiography}[{\includegraphics[width=1in,height=1.25in,clip,keepaspectratio]{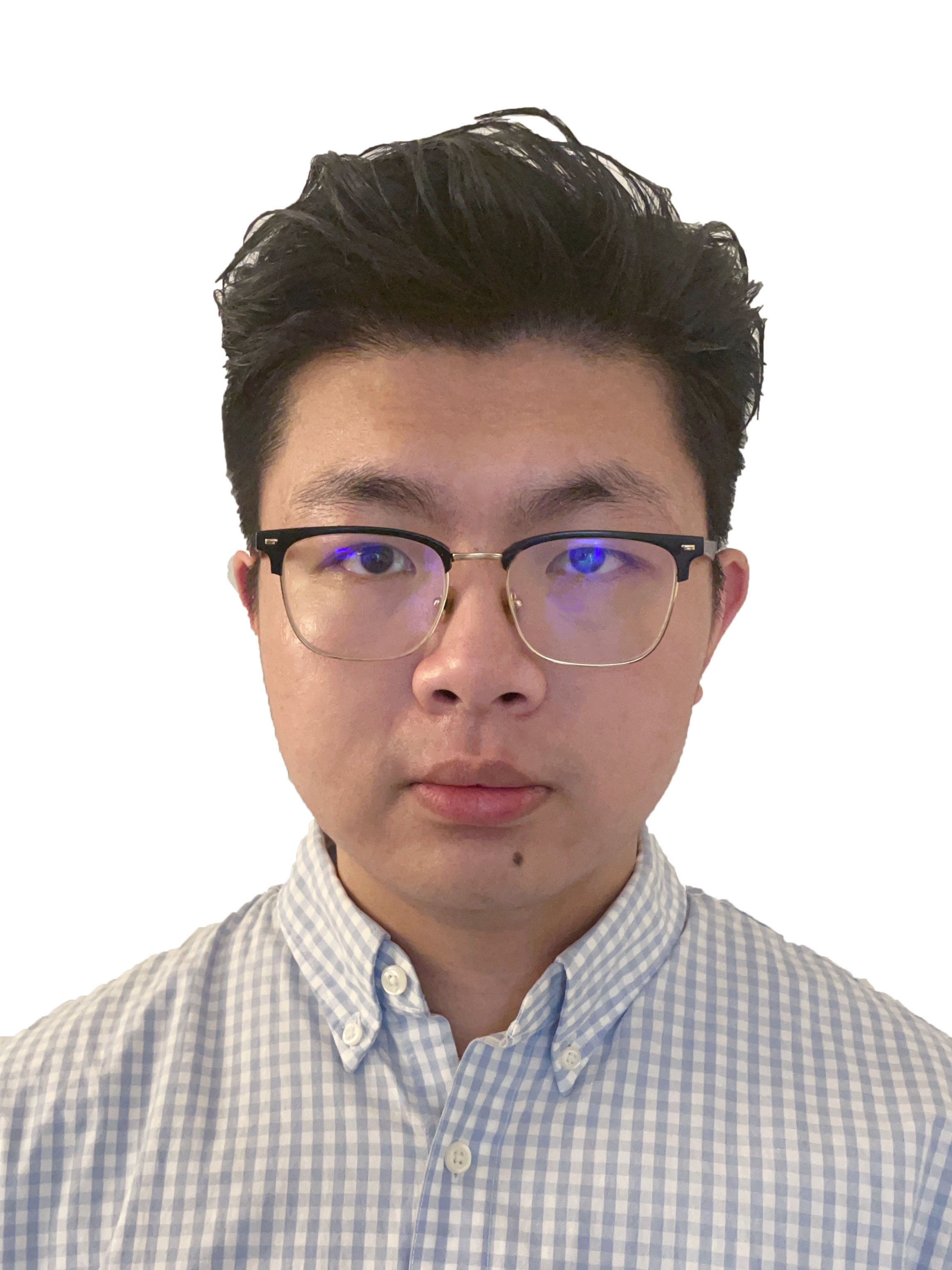}}]{Sicheng Mo} received his Bachelor's degree from the University of Wisconsin-Madison. He is pursuing a Master's degree at the University of California, Los Angeles. His research interest is computer vision.
\end{IEEEbiography}\vspace{-3em}

\begin{IEEEbiography}[{\includegraphics[width=1in,height=1.25in,clip,keepaspectratio]{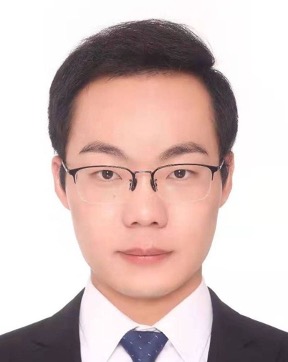}}]{Jiayong Peng} received his Bachelor's degree in Electronic Engineering from XiDian University, Xi'an, China in 2017. He is currently working towards a PhD degree at the University of Science and Technology of China. His research interests include computational photography, depth sensing and machine learning.
\end{IEEEbiography}\vspace{-3em}

\begin{IEEEbiography}[{\includegraphics[width=1in,height=1.25in,clip,keepaspectratio]{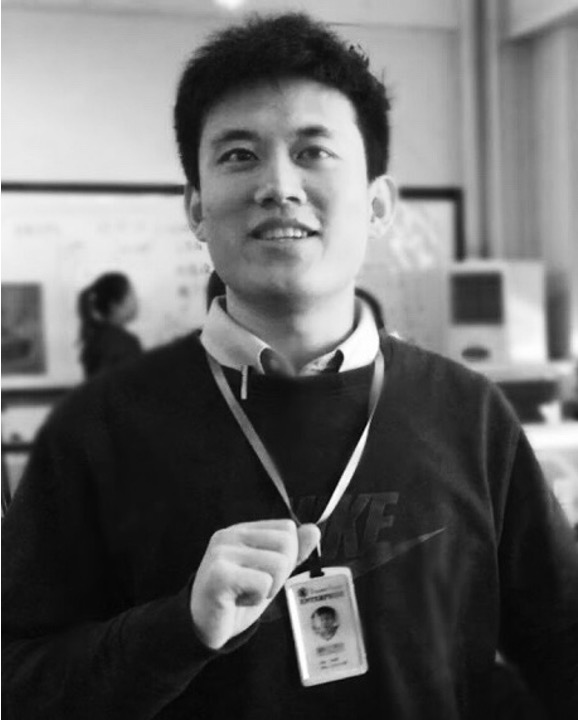}}]{Xiaochun Liu} received his PhD in Electrical and Computer Engineering from the University of Wisconsin-Madison. He joined Apple in 2022. His research interest is computational imaging.
\end{IEEEbiography}\vspace{-3em}

\begin{IEEEbiography}[{\includegraphics[width=1in,height=1.25in,clip,keepaspectratio]{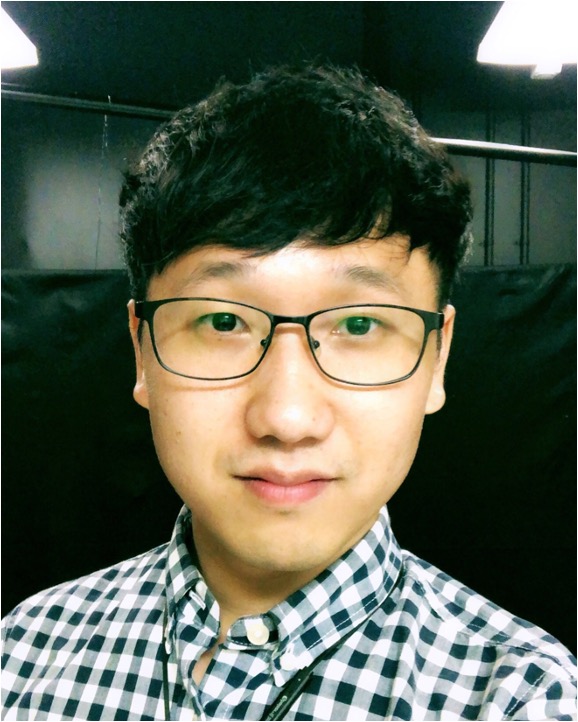}}]{Ji Hyun Nam} received his PhD in Electrical and Computer Engineering from the University of Wisconsin Madison with Professor Andreas Velten. His research focuses on computational imaging, machine learning, signal processing, light simulations and optics. He enjoys prototyping and engineering imaging systems and bringing them into the real-world. He developed computational algorithms and novel optical imaging systems that can see objects and scenes that are not directly visible. His representative works are published in Nature and Nature Communications. He received his Master's and Bachelor's degrees in Optical Engineering at Bauman Moscow State Technical University in 2016 and 2014.
\end{IEEEbiography}\vspace{-3em}

\begin{IEEEbiography}[{\includegraphics[width=1in,height=1.25in,clip,keepaspectratio]{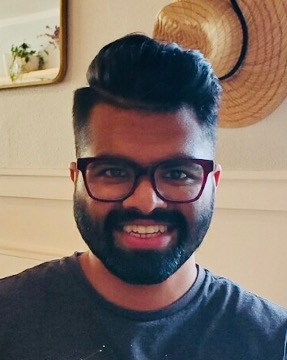}}]{Siddeshwar Raghavan} is a first-year ECE PhD student at Purdue University, advised by Professors Edward J. Delp and Fengqing Maggie Zhu. His research interests span across computer vision and machine learning. He graduated with a Master's degree from the ECE Department at the University of Wisconsin-Madison, where he worked with Professor Yin Li and Andreas Velten.
\end{IEEEbiography}\vspace{-3em}

\begin{IEEEbiography}[{\includegraphics[width=1in,height=1.25in,clip,keepaspectratio]{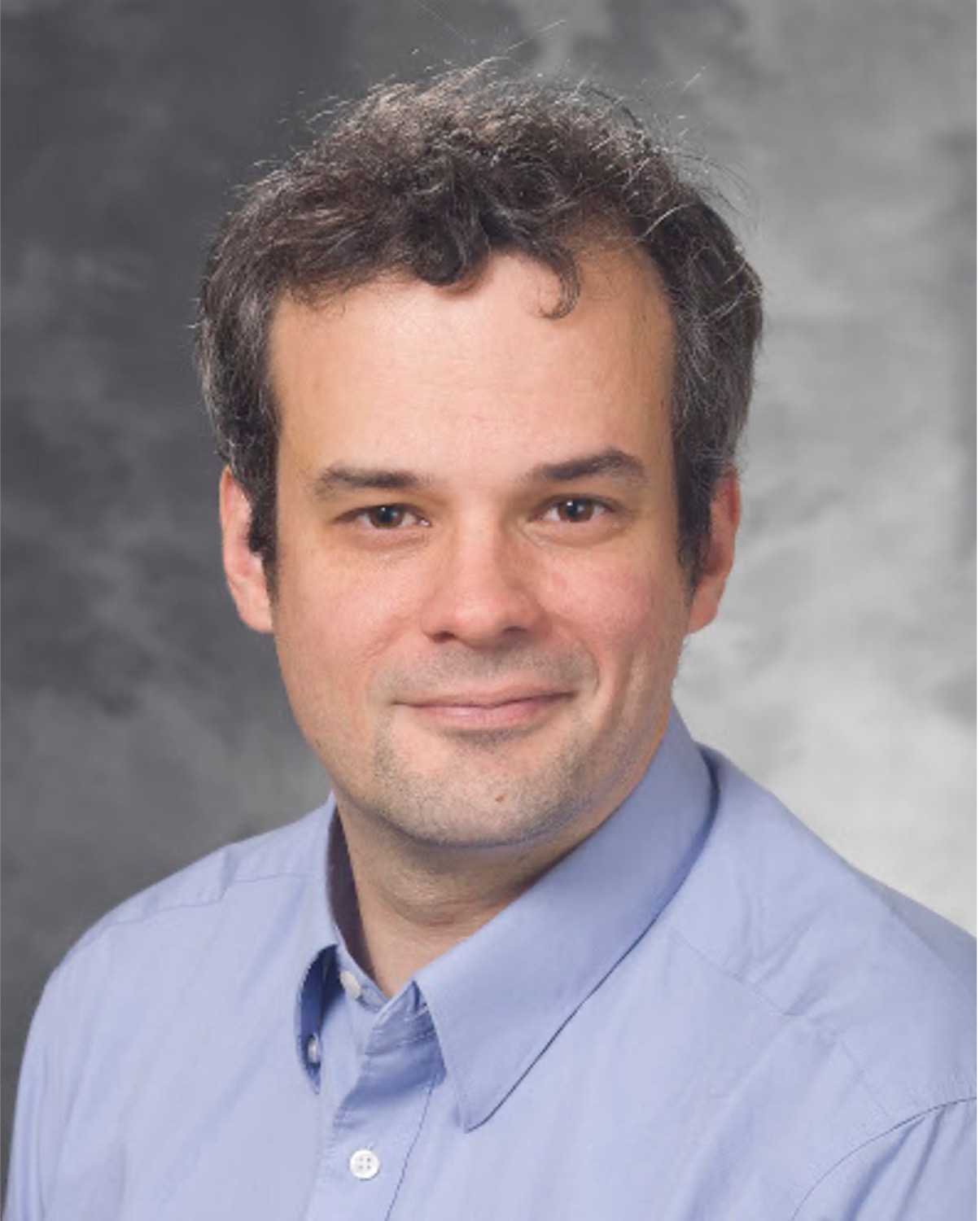}}]{Andreas Velten} received the BA degree in physics from the Julius Maximilian University of Wurzburg, Wurzburg, Germany in 2003, and the PhD degree in physics from the University of New Mexico, Albuquerque, NM, in 2009. From 2010 to 2012, he was a postdoctoral research associate with the Massachusetts Institute of Technology working on high speed imaging systems that can look around a corner using scattered light. From 2013 to 2016, he was an associate scientist with the Laboratory for Optical and Computational Instrumentation, University of Wisconsin–Madison, working in optics, computational imaging, and medical devices. Since 2016, he is an assistant professor with the Biostatistics and Medical Informatics, Electrical and Computer Engineering Department, University of Wisconsin–Madison. His research focuses on performing multidisciplinary work in applied computational optics and imaging.
\end{IEEEbiography}\vspace{-3em}

\begin{IEEEbiography}[{\includegraphics[width=1in,height=1.25in,clip,keepaspectratio]{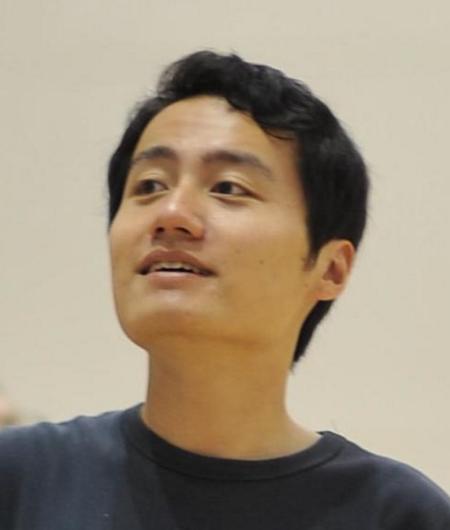}}]{Yin Li} is an Assistant Professor in the Department of Biostatistics and Medical Informatics and affiliate faculty in the Department of Computer Sciences at the University of Wisconsin-Madison. Previously, he obtained his PhD from the Georgia Institute of Technology and was a Postdoctoral Fellow at the Carnegie Mellon University. His primary research focus is computer vision. He is also interested in the applications of vision and learning for mobile health. Specifically, his group develops methods and systems to automatically analyze human activities for healthcare applications.
\end{IEEEbiography}
% \begin{IEEEbiography}[{\includegraphics[width=1in,height=1.25in,clip,keepaspectratio]{author_bio/yl.jpeg}}]{Yin Li}
% \end{IEEEbiography}


% \end{IEEEbiography}

%\newpage


%\vfill

\fi

% \section{Supplemental Materials}
% \begin{figure}[h]
% \centering
% \includegraphics[width=0.9\linewidth]{figures/ch1_128_sy_depth.pdf} 
% \caption{\textbf{Additional depth estimation results on the alphanumerics dataset.}}
% \label{unsup_netf}
% \end{figure}

% \begin{figure}[h]
% \centering
% \includegraphics[width=0.9\linewidth]{figures/n_views_2.pdf} 
% \caption{\textbf{Additional novel view synthesis results on a real measurement of digit ``2".}}
% \label{n_views_4}
% \end{figure}

\end{document}

%% file: tex/01_intro.tex
% The first section title should be wrapped inside a \IEEEraisesectionheading as follows.
\IEEEraisesectionheading{
  \section{Introduction}\label{sec:introduction}
}

% The very first letter of the paper is a 2 line initial drop letter
% followed by the rest of the first word in caps.
\IEEEPARstart{T}{ime-resolved} non-line-of-sight (NLOS) imaging recovers information about hidden scenes based on indirect reflectance scattered by the surrounding environment (e.g., a relay wall), and has the potential to revolutionize many critical applications such as medicine, robotics, military and law enforcement operations, and scientific imaging. A main limiting factor of existing NLOS imaging systems is the acquisition time, ranging from tens of seconds~\cite{liu2020phasor} to a few minutes~\cite{lindell2019wave,o2018confocal}. Recently, non-confocal NLOS imaging that combines a single laser scanning a relay wall and a sensor array capturing multiple transient measurements has emerged for rapid data acquisition, achieving an unprecedented frame rate up to 5Hz~\cite{nam2021low}. 

Such non-confocal NLOS imaging introduces approximations in light transport during data acquisition, most notably pixel remapping, in exchange for imaging speed, thereby posing new challenges for reconstructing the hidden scene using its transient measurements. Methods built on precise modeling of light paths~\cite{tsai2019beyond,shen2021non} will inevitably fall short due to inaccurate physics in this setup. Recent development has thus turned to the modeling of wave propagation~\cite{liu2020phasor,jiang2021ring}, resulting in methods that can successfully handle different surface reflectance and occlusions. Despite their robustness, these methods oftentimes yield reconstructions that lack fine details (e.g., textures and edges), and do not incorporate any prior about the hidden scenes.

A promising solution that has demonstrated recent success, mostly in confocal setups, is to integrate physical models into learning based approaches~\cite{shen2021non,chen2020learned}, such as deep neural networks. These methods enjoy the accuracy and robustness by modeling the underlying physics of NLOS imaging, while leveraging learned scene priors useful for reconstructing familiar shapes and visual details. Unfortunately, we observe that these prior approaches produce unsatisfactory results in a non-confocal setup, again, due to the approximations in physics necessary for rapid capturing. %in the non-confocal setup.
% Yin: NeTF uses physics based rendering that relies on precise modeling of light paths, and thus don't quite work. LFE, however, mostly uses neural networks for rendering without a strong assumption about the physics. See if you have better arguments here. 

To bridge this gap, we propose to embed physics models, consisting of an inverse propagation module and a volume renderer, into a deep neural network for non-confocal NLOS reconstruction. Specifically, our inverse propagation module adapts the Rayleigh-Sommerfeld diffraction (RSD) operator previously used in modeling wave propagation for NLOS~\cite{liu2020phasor}, and our physics-based volume renderer is inspired by neural transient field~\cite{shen2021non}. Our intuition is that using inverse propagation helps to regularize the solution space of the volume renderer, alleviating dependency on accurate modeling of light transport, thereby leading to robust generalization beyond an ideal image formation model. 

\fangzhou{Further, we devise a unified learning framework that enables our model to be flexibly trained using diverse supervision signals, including target intensity images or even raw NLOS transient measurements. Once trained, our model renders both intensity and depth images at inference time in a single forward pass with high efficiency. Through extensive experiments, we demonstrate that this orchestrated design yields superior reconstruction quality for both synthetic data and real captures.} %outperforming previous methods.

\fangzhou{Finally, we showcase several benefits of our method. \textit{First}, our model, despite being trained on synthetic data, generalizes well on real captures. \textit{Second}, our method, when implemented on a high-end GPU, processes 11.8 frames per second (FPS), thus paving the road towards fast NLOS imaging. \textit{Finally}, our method supports key functions beyond NLOS reconstruction; our model can synthesize images from a non-frontal view, and our learned features can facilitate accurate object recognition in NLOS imaging.}

%In addition to its high fidelity results, our model also celebrates several major benefits. \textit{First}, our model can be flexibly trained using various supervision signals including multi-view target images, or perhaps more importantly transient measurements themselves. \textit{Second}, once trained, our model can be used to render outputs beyond the training signals. For example, our model can be trained using intensity images or transient measurements, yet outputs depth maps at inference time. Further, our model is able to synthesis images from a non-frontal view. \textit{Third}, our method requires only a single forward pass for NLOS reconstruction, which can be efficiently implemented on a GPU, paving the road towards fast NLOS imaging and reconstruction. \textit{Fourth}, our method, despite being trained on synthetic data, generalizes well on real captures. \textit{Fifth}, the learned features from our model can be used for accurate object recognition in NLOS imaging.

\smallskip
\noindent \textbf{Scope and Limitation}: Our method is designed for non-confocal NLOS imaging that trades acquisition speed with approximations in light transport. While our method can be readily adapted to a confocal setup with precise physics, we conjecture that it is less likely to bring additional benefits to existing solutions such as~\cite{shen2021non} and~\cite{chen2020learned} in this setup. Further, there are fundamental limitations in NLOS reconstruction that are not addressed in our method, such as the challenging problem of missing cones~\cite{liu2019analysis}, despite some promising empirical results in our experiments. Finally, our method, similar to Nam et al.~\cite{nam2021low}, requires a high-end GPU to match the frame rate of the imaging system (5Hz) at inference time, which could be further optimized.

%% file: tex/02_related_works.tex
\section{Related Work}

\subsection{Non-Line-of-Sight Imaging}

Non-line-of-sight (NLOS) imaging aims at recovering the properties of a target scene hidden beyond the direct line of a camera's sight. With many applications including object detection~\cite{scheiner2020seeing,chen2020learned}, tracking~\cite{scheiner2020seeing,smith2018tracking} and human pose estimation~\cite{isogawa2020optical}, NLOS imaging has received considerable attention recently. Several imaging systems~\cite{velten2012recovering,lindell2019wave,musarra20193d,nam2021low} have been developed to capture multi-bounce indirect reflections from the hidden scene, scattered by the surrounding environment such as a diffuse relay surface. Accompanying with these systems are NLOS reconstruction methods that recover an ``image'' of the hidden scene. Most relevant to our work are methods that reveal the appearance and/or geometry of hidden objects using active illumination and time-resolved sensors, hence the focus of this section. A recent review on NLOS imaging can be found in~\cite{faccio2020non}.

Since the seminal work of Kirmani et al.~\cite{kirmani2009looking}, physics based NLOS reconstruction has seen rapid progress with major milestones falling into four categories, namely back-projection methods~\cite{velten2012recovering,o2018confocal}, wave propagation based methods~\cite{lindell2019wave,liu2019non, liu2020phasor, nam2021low}, iterative optimization methods~\cite{la2018error,tsai2019beyond,iseringhausen2020non} and geometry based methods~\cite{tsai2017geometry,xin2019theory}. Our method draws insight from wave based methods by adapting the inverse operator of Rayleigh-Sommerfeld diffraction (RSD)~\cite{liu2019non} for feature propagation, yet develops a learning based approach for NLOS reconstruction. Our learning objective is closely related to iterative optimization methods in which an update to a parametric object representation is issued via differentiable render-and-compare.

On the practical side, efficient hardware~\cite{nam2021low,liao2021fpga} and software implementations~\cite{arellano2017fast,liu2020phasor,jiang2021ring} have been developed for real-time NLOS imaging and reconstruction. Our method is tailored for the low-latency imaging hardware of Nam et al.~\cite{nam2021low} and adopts the fast RSD implementation of Liu et al.~\cite{liu2020phasor}. On the theoretical side, prior work has analyzed the feature visibility in NLOS measurements and concluded that hidden objects positioned in certain poses are inherently not detectable from a physics point of view~\cite{liu2019analysis}.
% characterized the role of Wigner distribution in RSD wave propagation~\cite{liu2020role}, provided justification to filtered back-projection through the lens of convolution~\cite{ahn2019convolutional}, and derived virtual light transport matrices for NLOS imaging~\cite{marco2021virtual}. 
We take a learning based approach and probe into the theoretical limit of NLOS reconstruction by learning statistical scene priors from large datasets.

\subsection{Learning-based NLOS Reconstruction}

Recently, learning based methods have emerged for NLOS reconstruction. Chopite et al.~\cite{chopite2020deep} presented the first deep model for NLOS reconstruction by training a U-Net~\cite{ronneberger2015u} on synthetic depth maps for depth estimation from transients. The fidelity of their results, however, falls short on real captures when compared with physics based methods, especially in the more challenging non-confocal imaging setup. More recent efforts incorporate physics based methods with learning based approaches. For example, Chen et al.~\cite{chen2020learned} proposed an efficient transient rasterizer and learned feature embeddings (LFE) on synthetic datasets for NLOS reconstruction and recognition. Shen et al.~\cite{shen2021non} introduced neural transient field (NeTF) for the implicit modeling of hidden objects. Zhu et al.~\cite{zhu2021fast} developed a deep generative model for NLOS imaging using inexpensive commercial LiDAR.

Our method is inspired by LFE and NeTF, yet with major advances. Both LFE and our method rely on a physics based inverse operator for feature propagation. However, LFE maps features to 2D images using a convolutional network, whereas our method volume-renders a radiance field conditioned on the features, leading to high fidelity reconstructions. NeTF is built around a volume renderer similar to ours. The main difference is that our method formulates reconstruction as feed-forward rendering, obviating the need for lengthy per-scene optimization, a key limitation of NeTF. From a learning perspective, both LFE and NeTF demand specific target (multi-view images and transients) for training, whereas our method, as we will demonstrate in the experiment, is more flexible in terms of supervision signals. Further, LFE and NeTF primarily consider a confocal NLOS setup, while our method targets the more challenging non-confocal setup with significantly higher acquisition speed.

\subsection{Neural Radiance Field}

Our method is also closely related to neural radiance field or NeRF, an implicit scene representation parametrized by a neural network that is initially designed for novel view synthesis~\cite{mildenhall2020nerf}. Numerous variants of NeRF have since been developed for various applications including video content generation~\cite{li2021neural}, object editing~\cite{liu2021editing}, and 3D-aware image synthesis~\cite{chan2021pi}. NeRF has also been combined with time-of-flight imaging for dynamic scene reconstruction~\cite{attal2021torf}. While most NeRF variants require iterative optimization on individual scenes, several conditioning mechanisms have been introduced for the sharing a template NeRF that can be adaptively modulated for feed-forward rendering of a scene. Prior work has conditioned NeRFs on random vectors for image generation~\cite{chan2021pi}, on learned embeddings for object editing~\cite{liu2021editing}, and on 2D image features for image-based rendering~\cite{wang2021ibrnet} and single-view 3D reconstruction~\cite{yu2021pixelnerf}. Our method is inspired by this line of work and for the first time learns a conditional radiance field for NLOS reconstruction.

%% file: tex/03_problem_statement.tex
\begin{figure}[!t]
\centering
\includegraphics[width=0.9\linewidth]{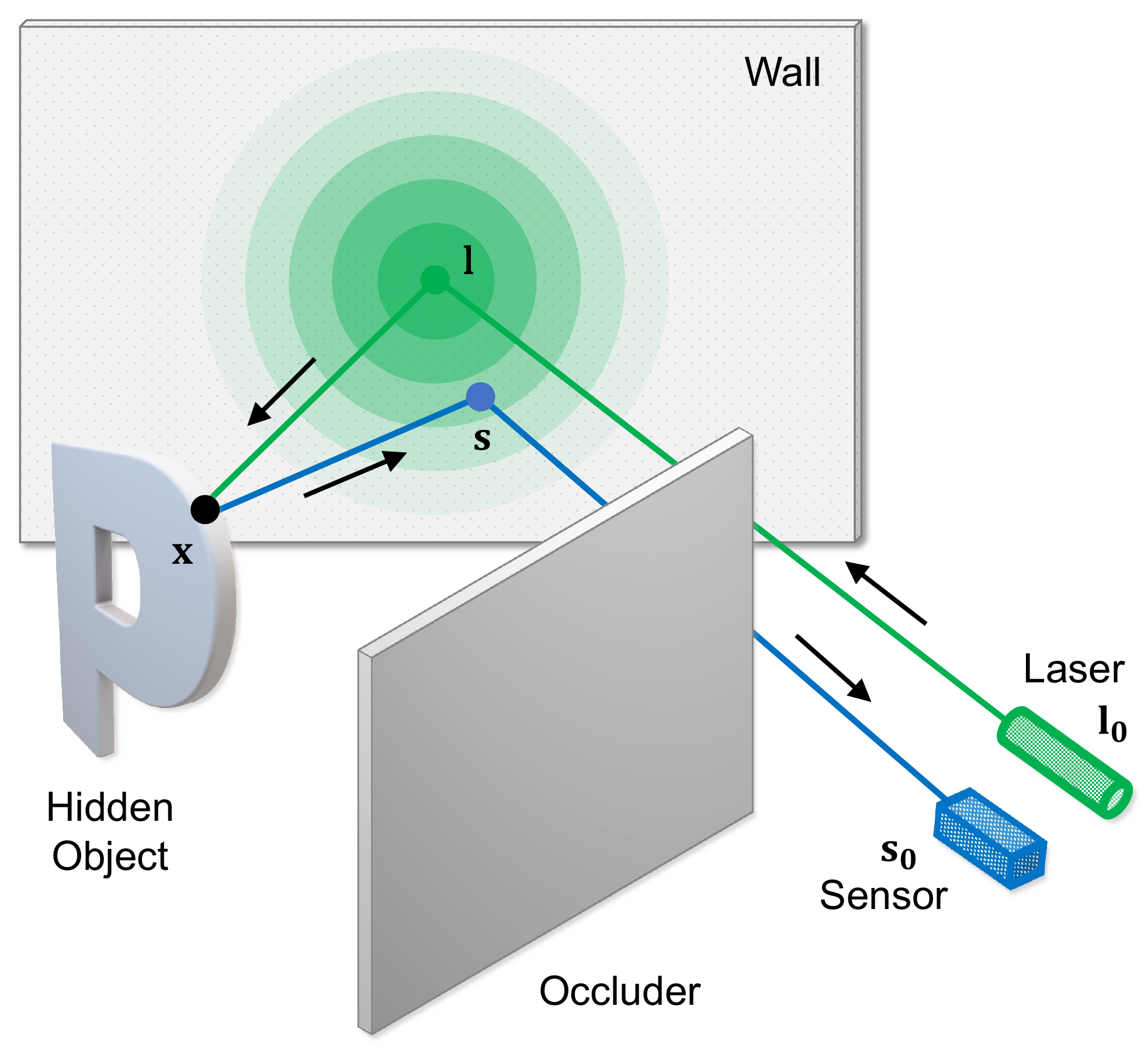} \vspace{-0.5em}
\caption{\textbf{Non-confocal NLOS imaging.} We illustrate a typical setup for non-confocal NLOS imaging. A pulsed laser located at $\mathbf{l}_0$ illuminates a part of a relay wall. Light bounces off the wall, interacts with the occluded object, scatters to the wall again, and is eventually captured by a time-resolved sensor at position $\mathbf{s}_0$ (different from $\mathbf{l}_0$).}
\label{nlos_figure}
\end{figure}

\section{Problem Statement}
We take a learning based approach to address the challenging task of non-line-of-sight (NLOS) imaging --- computationally reconstructing an object that is hidden around the corner. In this section, we present the image formation model for NLOS transient measurements and discuss the challenges our method aims to solve.
 
\subsection{Image Formation Model}
We assume an active, time-of-flight imaging system with a pulsed, collimated laser source at $\mathbf{l}_0$ and a transient sensor at $\mathbf{s}_0$. We further assume a diffuse, planar relay wall visible from both $\mathbf{l}_0$ and $\mathbf{s}_0$. As in Figure~\ref{nlos_figure}, a laser pulse emitted from $\mathbf{l}_0$ illuminates a wall location $\mathbf{l}$, where light scatters off the wall and travels towards the hidden object. Light reaching a point $\mathbf{x}$ on the object gets reflected once again and heads back to the wall. Finally, returning photons that hit wall location $\mathbf{s}$ are recorded by the sensor pointing at $\mathbf{s}$~\footnote{We do not model higher-order bounces for simplicity, acknowledging that they also contribute to sensor response and can be valuable to reconstruction~\cite{marco2021virtual}.}.

If one regards $\mathbf{l}$ and $\mathbf{s}$ respectively as \emph{virtual} laser source and sensor, light transport along the three-bounce path $\mathbf{l}_0 \rightarrow \mathbf{l} \rightarrow \mathbf{x} \rightarrow \mathbf{s} \rightarrow \mathbf{s}_0$ conceptually simplifies to direct illumination $\mathbf{l} \rightarrow \mathbf{x} \rightarrow \mathbf{s}$. Mathematically, the impulse response $\tau_{(\mathbf{l},\mathbf{s})}(t)$ can be written in path integral form as
\begin{equation}\label{theory}
\begin{split}
    \tau_{(\mathbf{l},\mathbf{s})}(t)=\int_{\mathcal{A}}\ &f_r(\mathbf{n}_{\mathbf{x}},\omega_{\mathbf{x}\rightarrow\mathbf{l}},\omega_{\mathbf{x}\rightarrow\mathbf{s}})G(\mathbf{l}\leftrightarrow\mathbf{x})G(\mathbf{s}\leftrightarrow\mathbf{x})\\&\delta(\|\mathbf{l}-\mathbf{x}\|_2+\|\mathbf{s}-\mathbf{x}\|_2-ct)\ \mathrm{d}A(\mathbf{x}),
\end{split}
\end{equation}
where $\mathrm{d}A(\mathbf{x})$ is a small surface of the hidden object surrounding $\mathbf{x}$, and $\mathcal{A}$ is the union of all surfaces. $f_r$ is the bi-directional reflectance distribution function (BRDF), $\mathbf{n}_{\mathbf{x}}$ the surface normal at $\mathbf{x}$, $\omega_{\cdot\rightarrow\cdot}$ a unit directional vector, $\delta$ the Dirac delta, $c$ the speed of light, and 
\begin{equation}
    G(\mathbf{l}\leftrightarrow\mathbf{x})=\frac{\langle\omega_{\mathbf{x}\rightarrow\mathbf{l}},\mathbf{n}_{\mathbf{x}}\rangle}{\|\mathbf{l}-\mathbf{x}\|_2^2}\cdot V(\mathbf{l}\leftrightarrow\mathbf{x}),
\end{equation}
\begin{equation}
    G(\mathbf{s}\leftrightarrow\mathbf{x})=\frac{\langle\omega_{\mathbf{x}\rightarrow\mathbf{s}},\mathbf{n}_{\mathbf{x}}\rangle}{\|\mathbf{s}-\mathbf{x}\|_2^2}\cdot V(\mathbf{s}\leftrightarrow\mathbf{x}).
\end{equation}
$V(\cdot\leftrightarrow\cdot)$ is the visibility function that evaluates to 1 if the two points are visible to each other.

Without loss of generality, we assume the sensor is a single photon avalanche diode (SPAD) with bin resolution $\Delta t$. Under suitable laser power and the sensor model of Hernandez et al.~\cite{Hernandez2017SPAD}, $\tau_{(\mathbf{l},\mathbf{s})}(t)$ over the time range $[0, T\Delta t)$ yields a transient histogram of photon counts $\mathbf{h}_{(\mathbf{l},\mathbf{s})}$ with $T$ temporal bins. This represents the ultimate sensor output that constitutes an NLOS measurement.

\subsection{Non-Confocal vs. Confocal Measurements}

An NLOS measurement is a 2D array of transient histograms $\{\mathbf{h}_{(\mathbf{l}_{ij},\mathbf{s}_{ij})}\}_{i=1...H,j=1...W}$ sampled at a dense $H \times W$ grid of locations $\mathbf{s}_{ij}$ on the wall. Importantly, a \emph{non-confocal} measurement assumes a fixed virtual laser source located at the wall center $\mathbf{l}$, that is, $\forall i,j$, $\mathbf{l}_{ij}=\mathbf{l}$, whereas in the traditional \emph{confocal} setting, the virtual sensor and laser source always co-localize, that is, $\forall i,j$, $\mathbf{l}_{ij}=\mathbf{s}_{ij}$. Most research to date has focused on confocal reconstruction owing to its simpler physics, leaving the more challenging non-confocal reconstruction problem less explored.

One key advantage of the non-confocal setup is its rapid data acquisition speed. Experimentally, a confocal measurement is acquired by a dense, exhaustive scan of the wall, which takes tens of seconds to complete for the existing hardware~\cite{lindell2019wave}. In contrast, the state-of-the-art non-confocal system runs at 5 FPS~\cite{nam2021low}, enabling numerous applications that require high-speed imaging. This is achieved by exploiting two key ideas: (1) reciprocity of light transport; (2) a sparse scan pattern with local pixel remapping. \fangzhou{In particular, pixel remapping enables the collection of \emph{dense} measurements with a \emph{sparse} raster scan of the relay wall. This is achieved by recording multiple responses simultaneously using a pixel array (vs. a single pixel in confocal systems) and mapping pixels to skipped positions between two scanlines subject to some approximation in geometry~\cite{nam2021low}.}

In this paper, we assume \emph{the non-confocal setup}, which we believe represents a more promising direction for future development of NLOS imaging systems.

\subsection{Key Challenges}

\fangzhou{Practically, the non-confocal setup poses major challenges for NLOS reconstruction owing to the numerous approximations introduced by the light transport model and hardware implementation.
% With numerous approximations in light transport, the non-confocal setup poses major challenges for NLOS reconstruction. 
Rather than a physics based method, we focus on \emph{a learning based} approach for non-confocal NLOS reconstruction for its potential to accommodate the approximations while recovering fine image details.}

A key obstacle to learning based NLOS reconstruction is the lack of a large dataset of \emph{real} measurements for training. Existing models trained on synthetic data generalize poorly on real data. This is because the synthetic data respect an ideal image formation model (Equation~\ref{theory}), yet a practical imaging system inevitably introduces approximations including calibration error, unmodeled sensor noise, imperfect wall geometry and reflectance, among others. Existing synthetic datasets also ignore potentially important aspects of light transport, such as multiple scattering, subsurface scattering, complex surface BRDF and fine surface geometry details. To make the situation worse, the fast non-confocal system makes additional approximations, most notably pixel remapping, in exchange for imaging speed (Figure~\ref{real_fig}(a)), further breaking existing learning based solutions. This raises the first challenge:

\begin{tcolorbox}%[colback=white, title=\textbf{Challenge 1}]
  \textbf{Challenge 1:} 
  Can we design a \emph{robust} learning based model that generalizes well on real measurements despite being trained on synthetic data?
\end{tcolorbox}

On the other hand, existing learning based models have separately explored different supervision targets including multi-view images~\cite{chen2020learned}, depth maps~\cite{chopite2020deep} and transients themselves~\cite{shen2021non}. However, a model designed with one target in mind can neither be trained on nor be tasked to generate another target. This limits the model's use case and prevents it from learning a holistic representation for best reconstruction results, hence our second challenge:

\begin{tcolorbox}%[colback=white, title=\textbf{Challenge 2}]
  \textbf{Challenge 2:} 
  Can we design a \emph{unified} learning framework that makes use of diverse training signals?
\end{tcolorbox}

%% file: tex/04_method.tex
\section{Deep NLOS Reconstruction with Physics Priors}

We propose a robust deep model for NLOS imaging tailored for the latest non-confocal system that runs at a high frame rate. 
% Our key innovation is to incorporate strong physics prior throughout the model to combat approximations in light transport during data acquisition made in exchange for imaging speed.
Our key innovation towards addressing \textbf{Challenge 1} is to incorporate the complementary physics priors of wave propagation and volume rendering into our deep model. We hypothesize that this orchestrated design encourages a model trained on synthetic data to generalize well on high-speed real captures.% subject to numerous approximations in light transport.
%Further, we devise a learning framework that supports diverse supervision signals without modification to the model architecture. Trained on synthetic data, our method outperforms existing physics and learning based approaches on both synthetic and real measurements.

\subsection{Method Overview}

\begin{figure}[!t]
\centering
\includegraphics[width=\linewidth]{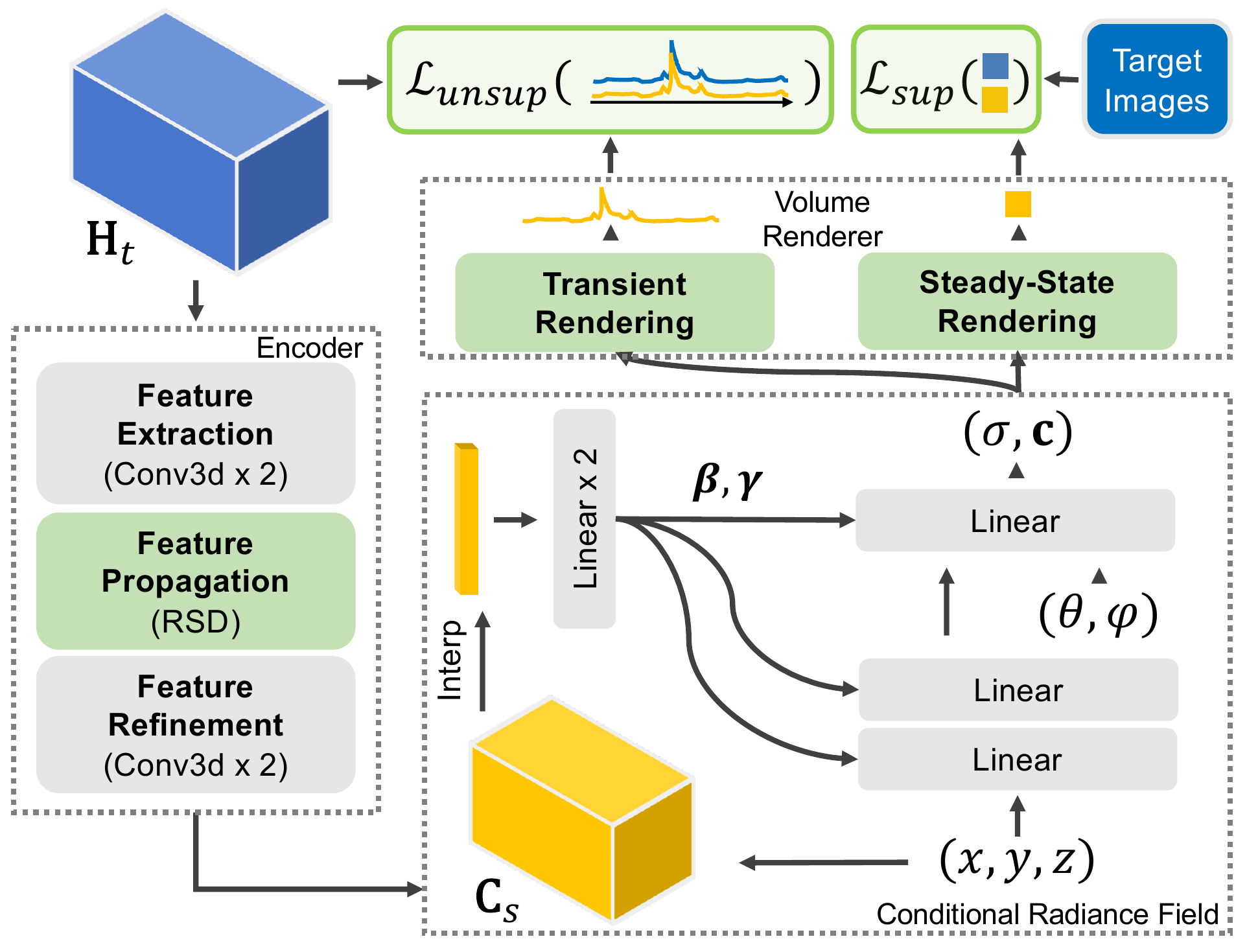} \vspace{-2em}
\caption{\textbf{Method Overview.} Our model consists of an encoder, a conditional radiance field and a volume renderer. Our volume renderer can flexibly synthesize 2D images using a steady-state forward model, or temporal histograms using a transient forward model at training time for different learning strategies. At inference time, our method renders 2D images in a feed-forward manner for NLOS reconstruction. {\color{LimeGreen}{\textbf{Green}}} blocks for physics-based, parameter-free modules, and \textcolor{gray}{gray} blocks for learnable modules.}
\label{method_figure}
\end{figure}

Figure~\ref{method_figure} presents an overview of our method. Our model consists of three components. First, an {\bf encoder} extracts features from the time-domain measurement and back-projects the features to the spatial domain via physics based inverse propagation (Section~\ref{encoder}). A {\bf neural radiance field} is subsequently conditioned on the spatial features to yield the shape and appearance of the hidden object (Section~\ref{crf}). This representation is further rendered into measurements of the desired sensor type for training and inference using a physics based {\bf volume renderer} (Section~\ref{renderer}).

In the absence of a large dataset of real measurements, we train our model on synthetic data. Our framework naturally supports two training strategies rooted in the idea of render-and-compare: (1) {\bf supervised learning} by comparing rendered and ground-truth 2D images; (2) {\bf unsupervised learning} by matching rendered histograms to the input measurement. We present their respective training objectives in Section~\ref{training}. Importantly, we render 2D images at inference time and interpret them as reconstruction of the hidden object.

\subsection{Encoder}
\label{encoder}

The encoder transforms the histograms of photon counts into time-domain features and propagates them to the spatial domain via a physics based inverse operator. 

\smallskip
\noindent \textbf{Feature Extraction}:
Given a 2D array of histograms $\mathbf{H}_t$ of size $H \times W \times T$, a strided 3D convolution with kernel size 3 immediately down-samples the input along all spatiotemporal axes to form a feature cube $\mathbf{C}_t$ of size $H/2 \times W/2 \times T/2 \times C$, where $C$ is the introduced feature dimension. The features are further processed by another 3D convolution with kernel size 3 before undergoing RSD inverse propagation as detailed next. Compared to the ResNet-like~\cite{he2016deep} design of Chen et al.~\cite{chen2020learned}, our feature extraction module is more parameter-efficient and facilitates stable training.

\smallskip
\noindent \textbf{Feature Propagation}:
\fangzhou{Physically, a non-confocal NLOS measurement represents the impulse response of the hidden scene to a pulsed signal emitted from $\mathbf{l}$. Liu et al.~\cite{liu2019non} showed that one may convolve it with any illumination wave $\mathcal{P}_{\mathbf{l}}(t)$ to simulate a \emph{virtual} line-of-sight (LOS) transient camera behind the wall and apply the LOS wave propagation theory of RSD for NLOS reconstruction.}
% Physically, back-projection of non-confocal NLOS measurements is governed by the RSD kernel~\cite{liu2019non}. 
Importantly, Nam et al.~\cite{nam2021low} observed that RSD is resilient to local rearrangement of sensor response, a key approximation that enables rapid NLOS imaging. 

\fangzhou{Drawing inspiration from the theory, we interpret $\mathbf{C}_t$ as the \emph{featurized} impulse response and leverage RSD as a robust physics prior to propagate $\mathbf{C}_t$ to the spatial domain}
% We hence leverage RSD as a robust physics prior to propagate time-domain features $\mathbf{C}_t$ to the spatial domain
\begin{equation}
    \mathbf{C}_s = \mathrm{RSD}(\mathcal{P}_{\mathbf{l}}(t)*\mathbf{C}_t),
\end{equation}
where $*$ is 1D convolution along the time axis and $\mathrm{RSD}(\cdot)$ is applied separately on individual feature channels. $\mathbf{C}_s$ is the spatial-domain feature of size $H/2 \times W/2 \times D \times C$ with $D$ the number of depth planes. Note that RSD is fully differentiable, hence enables end-to-end training of our model. We adopt the implementation of Liu et al.~\cite{liu2020phasor} for its low computational and memory cost and refer readers to~\cite{liu2020phasor} for mathematical and implementation details. % More details can be found in the appendix. 

\smallskip
\noindent \textbf{Feature Refinement}:
The RSD output at reduced resolution is prone to artifacts due to aliasing. The encoder thus further refines $\mathbf{C}_s$ using two 3D convolutions with kernel size 3. The refined features subsequently participate in rendering.

\subsection{Conditional Neural Radiance Field}
\label{crf}

Central to our method is a radiance field $F_{\Theta}: (\mathbf{x}, \mathbf{d}; \mathbf{C}_s)\rightarrow(\mathbf{c}, \sigma)$ that is \emph{ conditioned} on the encoded features $\mathbf{C}_s$. Similar to NeRF~\cite{mildenhall2020nerf}, $F$ is realized as a multi-layer perceptron (MLP) with weights $\Theta$ that maps 3D location $\mathbf{x}=(x,y,z)$ and 2D viewing direction $\mathbf{d}=(\theta,\phi)$, both position encoded, to intensity or color $\mathbf{c}$ and volume density $\sigma$. Unlike NeRF and its transient variant NeTF~\cite{shen2021non} that learn separate $\Theta$ for each individual scene, our representation shares $\Theta$ across all scenes, specializes via the adaptive modulation of network activations dependent on $\mathbf{C}_s$ (Figure~\ref{method_figure}), and thus removes the need for iterative optimization of $\Theta$ at inference time. 

Our key intuition is in three folds: (1) learning a shared template facilitates the distillation of scene priors from diverse training data; (2) conditioning on scene-dependent features enables fast \emph{feed-forward} reconstruction, bypassing lengthy per-scene training; and most importantly, (3) our representation seamlessly bridges the encoder and volume renderer, both physics based, thus placing strong mutual constraints on their learning. In doing so, our model, despite being trained exclusively on synthetic data, generalizes well on real measurements as we show in our experiments.

Our conditioning mechanism goes as follows. We sample $\mathbf{C}_s$ at $\mathbf{x}$ via tri-linear interpolation and feed it to a small MLP to predict the per-unit affine weight $\boldsymbol{\gamma}_i$ and bias $\boldsymbol{\beta}_i$ for the activations $\mathbf{h}_i$ at layer $i$. The activations are then transformed as
\begin{equation}
    \mathbf{h}'_i = \boldsymbol{\gamma}_i \odot \mathbf{h}_i + \boldsymbol{\beta}_i,
\end{equation}
where $\odot$ stands for element-wise multiplication. Note that $\mathbf{C}_s$ has injected scene-specific knowledge into the activations $\mathbf{h}'_i$ through the affine parameters. Our approach is inspired by $\pi$-GAN~\cite{chan2021pi}, ECRF~\cite{liu2021editing} and pixelNeRF~\cite{yu2021pixelnerf}, which similarly condition a radiance field on latent vectors for image generation, object editing and novel view synthesis.

\subsection{Volume Rendering}
\label{renderer}

One key strength of our physics based volume rendering framework is its capacity to render the conditional radiance field $F$ into measurements of any sensor type given an appropriate forward rendering model. Our method explores two forward models for training and inference, namely the steady-state rendering of 2D images and transient rendering of temporal histograms. Importantly, our forward models are fully differentiable, hence enables the end-to-end training of our deep model. We focus on their mathematical formulation in this section and discuss their connection with the training objectives in the next section.

\smallskip
\noindent \textbf{Steady-State Rendering}: We adopt the same forward model as NeRF~\cite{mildenhall2020nerf} for the steady-state rendering of $F$ into 2D images. Briefly, the rendered pixel value $\hat{I}(\mathbf{r})$ for a ray $\mathbf{r}(u) = \mathbf{o} + u\mathbf{d}$ originating from camera center $\mathbf{o}$ along viewing direction $\mathbf{d}$ is
\begin{equation}\label{ss_img}
    \hat{I}(\mathbf{r}) = \int_0^\infty T(u)\ \sigma(\mathbf{r}(u))\mathbf{c}(\mathbf{r}(u), \mathbf{d})\ \mathrm{d}u,
\end{equation}
where $T(u) = \exp\left(-\int_0^u\sigma(\mathbf{r}(s))\mathrm{d}s\right)$ is the cumulative transmittance over distance $u$. We bound the integration range and numerically estimate the integral as in~\cite{mildenhall2020nerf}. 

\fangzhou{A similar formulation can be used to render a depth map from $F$. Specifically, the expected distance $\hat{d}$ of a ray $\mathbf{r}$ colliding with a scene point is given by 
\begin{equation}\label{ss_dep}
    \hat{d}(\mathbf{r}) = \int_0^\infty T(u)\ \sigma(\mathbf{r}(u))u\ \mathrm{d}u.
\end{equation}}

\begin{figure}[!t]
\centering
\includegraphics[width=0.9\linewidth]{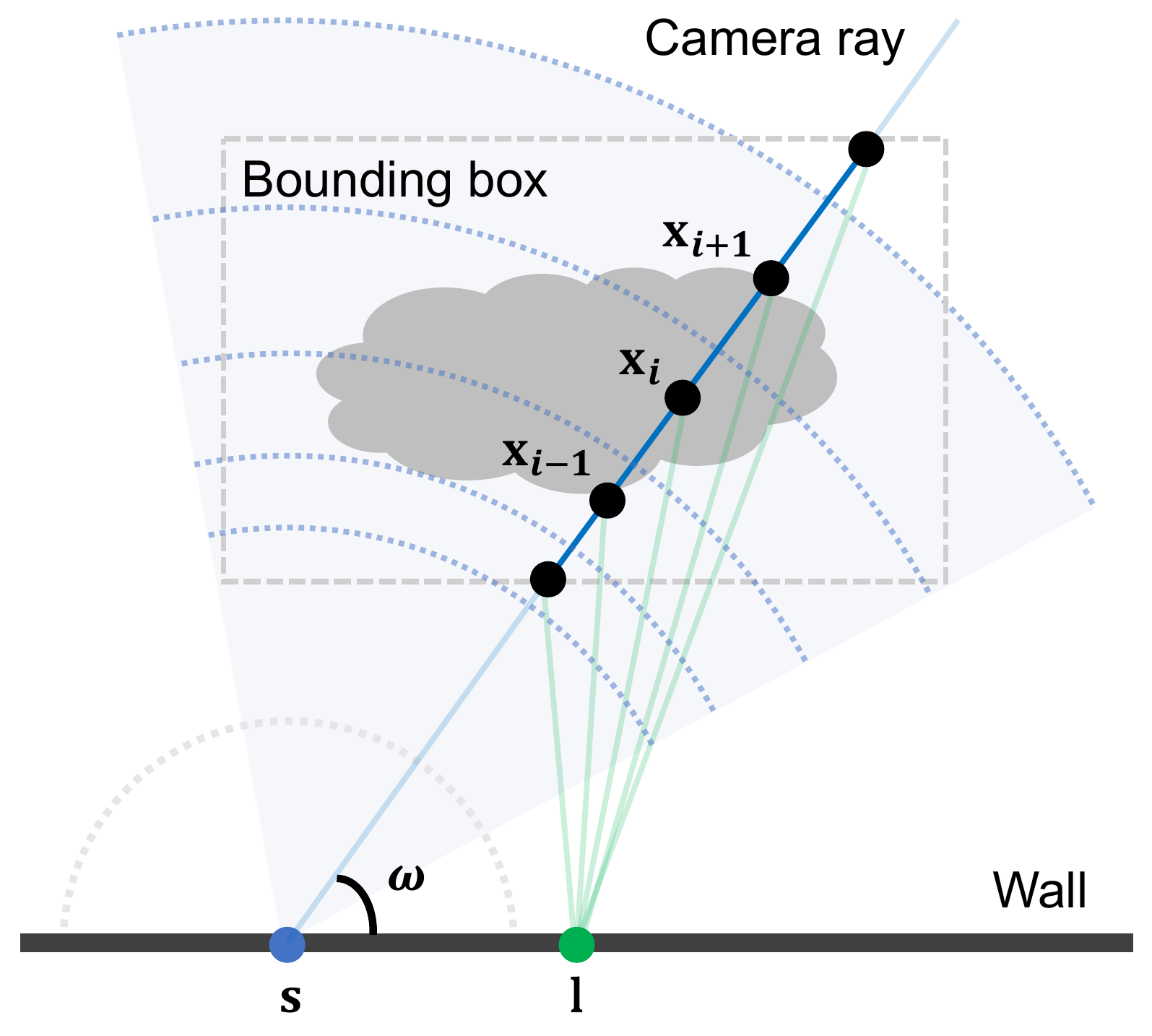} \vspace{-1em}
\caption{\fangzhou{\textbf{Ray and point sampling for transient rendering.} Rays originating from a virtual sensor $\mathbf{s}$ are uniformly drawn from within the cone shaded in blue so that they always intersect the object bounding box. Scene points $\mathbf{x_i}$ are sampled along a ray such that the length of path $\mathbf{l}\rightarrow\mathbf{x_i}\rightarrow\mathbf{s}$ is uniformly distributed. Note that the length of camera subpath $\mathbf{s}\rightarrow\mathbf{x_i}$ is \emph{not} uniformly distributed, and no point is drawn from light subpath $\mathbf{l}\rightarrow\mathbf{x_i}$ as we only model outgoing radiance.}}
\label{sampling}
\end{figure}

\smallskip
\noindent \textbf{Transient Rendering}: In the non-confocal setting, the hidden object can be considered as being illuminated by a virtual, pulsed laser source located at the wall center $\mathbf{l}$. To facilitate the differentiable rendering of $F$ into a transient histogram with time resolution $\Delta t$ captured by a virtual sensor at wall location $\mathbf{s}$, we approximate the surface integral in Equation~\ref{theory} with a volume integral.
%to facilitate gradient backpropagation. 
Concretely, we calculate the Poisson rate $\hat{\tau}_i$ of the $i$-th bin by summing the contribution of all points $\mathbf{x}$ within a hemi-ellipsoidal shell such that light propagation time along the (reversed) path $\mathbf{s} \rightarrow \mathbf{x} \rightarrow \mathbf{l}$ falls within $\left[i\Delta t, (i+1)\Delta t \right)$. Formally,
\begin{equation}\label{eq1}
    \hat{\tau}_i = \int_{i\Delta t}^{(i+1)\Delta t}\ \hat{L}(t)\ \mathrm{d}t,
\end{equation}
\begin{equation}\label{eq2}
    \hat{L}(t) = \int_{\phi}\int_{\theta}\  T(u(t))\sigma(\mathbf{r}(u(t)))\mathbf{c}(\mathbf{r}(u(t)), \mathbf{d}) \ \mathrm{d}\theta\mathrm{d}\phi,
\end{equation}
where $u(t)=\frac{c^2t^2 - \left\| \mathbf{\mathbf{s}-\mathbf{l}} \right\|_2^2}{2ct - 2\cos{\theta}}$ is the distance from $\mathbf{s}$ to a point $\mathbf{x}$ along the ray direction $\mathbf{d}=(\theta,\phi)$ such that the path $\mathbf{s} \rightarrow \mathbf{x} \rightarrow \mathbf{l}$ has length $ct$. % Our formulation can be viewed as a differentiable approximation to the transient image formation model (Equation~\ref{theory}) in which the surface integral is substituted by a volume integral.

In practice, we interchange the order of integrals in Equations (\ref{eq1}) and (\ref{eq2}), and apply the line-to-point sampling strategy of Jarabo et al.~\cite{jarabo2014framework} for the \emph{unbiased} estimation of $\hat{\tau}_i$. Further, we assume an axis-aligned bounding box around the hidden object and adapt the method of Urena et al.~\cite{urena2013area} to draw ray directions from within the spherical projection of the box's proximal face relative to the wall. This encourages the ray samples and in turn the point samples along a ray to concentrate around the hidden object. Our sampling strategy is illustrated in Figure~\ref{sampling}.

\smallskip
\noindent \textbf{Comparison to NeTF~\cite{shen2021non}}:
Our transient rendering formulation differs from Shen et al.~\cite{shen2021non} in three aspects. First, we estimate the outgoing radiance at a scene point without factoring it into irradiance and BRDF. This simplification, in line with NeRF~\cite{mildenhall2020nerf}, allows the implicit modeling of occlusion and multi-bounce light and better supports feed-forward rendering. Second, we present a more principled and efficient sampling framework for per-bin Poisson rate estimation. Finally, we empirically found that our proposed conditional scene parametrization bootstraps the learning of $F$, and hence none of the advanced training techniques (e.g., two-stage training, hierarchical sampling) from~\cite{shen2021non} is needed. \fangzhou{Our experiments show that our transient rendering recipe can be trivially adapted for iterative optimization and yields better reconstruction than NeTF.}

%% file: tex/05_training_inference.tex
\section{Model Training and Inference}\label{training}

We now present the training and inference schemes of our method. Our key idea is to minimize a reconstruction loss between the rendered and ground-truth sensor measurements, often known as render-and-compare. Importantly, one favorable property of our framework towards addressing \textbf{Challenge 2} is its capacity to support flexible reconstruction objectives for supervised learning as well as unsupervised learning. \fangzhou{Similarly, our model readily supports the rendering of various targets at inference time including intensity and depth images.}

\subsection{Training Schemes}

\smallskip
\noindent \textbf{Supervised Training}: When multi-view target images are available in the training data, we minimize the mean squared error (MSE) between the rendered and ground-truth pixel values
\begin{equation}
    \mathcal{L}_{mse} = \frac{1}{|\mathcal{R}|} \sum_{\mathbf{r} \in \mathcal{R}} \left\| \hat{I}(\mathbf{r}) - I(\mathbf{r}) \right\|_2^2,
\end{equation}
where $\mathcal{R}$ is the set of sampled rays. Following Lombardi et al.~\cite{lombardi2019neural}, we add a Beta distribution prior on the cumulative transmittance $T_{\mathbf{r}}$ to bias a ray $\mathbf{r}$ towards either hitting or missing the object
\begin{equation}
    \mathcal{L}_{beta} = \frac{1}{|\mathcal{R}|} \sum_{\mathbf{r} \in \mathcal{R}} \log(T_{\mathbf{r}}) + \log(1 - T_{\mathbf{r}}),
\end{equation}
and a total variation (TV) prior on log opacities
\begin{equation}\label{tv_eq}
    \mathcal{L}_{tv} = \frac{1}{|\mathcal{R}|} \sum_{\mathbf{r} \in \mathcal{R}}\sum_{i}\|\fangzhou{\Delta\log\alpha(\mathbf{r})_i}\|_1,
\end{equation}
where $i$ indexes over the sampled steps along $\mathbf{r}$ and $\alpha(\mathbf{r})_i=1-\exp{\left(\sigma(\mathbf{r}(s_i))\Delta s_i\right)}$ is the opacity at $\mathbf{r}(s_i)$ over the discretized step $\Delta s_i$.

The full training objective is a weighted combination of the three terms
\begin{equation}
    \mathcal{L}_{sup} = \lambda_{mse}\mathcal{L}_{mse} + \lambda_{beta}\mathcal{L}_{beta} + \lambda_{tv}\mathcal{L}_{tv},
\end{equation}
with $\lambda_{mse}$=1, $\lambda_{beta}$=0.0001 and $\lambda_{tv}$=0.01 in our experiments.

\smallskip
\noindent \textbf{Unsupervised Training}: In the absence of target images, we minimize the Poisson negative log-likelihood~\footnote{We omit the constant term $\log(n_{\mathbf{s}, b}!)$ for conciseness.} of the rendered transient histograms
\begin{equation}
    \mathcal{L}_{poisson} = \frac{1}{|\mathcal{S}||\mathcal{B}|} \sum_{\mathcal{S}} \sum_{\mathcal{B}} \hat{\tau}_{\mathbf{s}, b} - n_{\mathbf{s}, b} \log(\hat{\tau}_{\mathbf{s}, b}),
\end{equation}
where $\hat{\tau}_{\mathbf{s}, b}$ is the rendered Poisson rate of the $b$-th bin at virtual sensor location $\mathbf{s}$, $n_{\mathbf{s}, b}$ the photon counts in the same bin of the target histogram, $\mathcal{B}$ the set of rendered bins and $\mathcal{S}$ the set of sampled sensor locations. The full objective is 
\begin{equation}
    \mathcal{L}_{unsup} = \lambda_{poisson}\mathcal{L}_{poisson} + \lambda_{beta}\mathcal{L}_{beta} + \lambda_{tv}\mathcal{L}_{tv},
\end{equation}
with $\lambda_{poisson}$=1, $\lambda_{beta}$=0.0001 and $\lambda_{tv}$=0.01.

\smallskip
\noindent \textbf{Joint Training}: One may combine the supervised and unsupervised learning objectives and arrive at what we can the joint training objective
\begin{equation}
    \mathcal{L}_{joint} = \mathcal{L}_{sup} + \mathcal{L}_{unsup}.
\end{equation}

\subsection{Inference}
\fangzhou{At inference time, we equip the volume renderer with the steady-state forward model (Equations (\ref{ss_img}) and (\ref{ss_dep})) to synthesize intensity and depth images of the hidden scene \emph{in a single forward pass}.}

\smallskip
% \noindent \textbf{Real-time Reconstruction}: We compare the inference time of different methods in Table~\ref{runtime}. Notably, our model runs at \emph{11.8 FPS} on an NVIDIA V100 GPU. This exceeds the data acquisition rate (5 FPS) of the high-speed imaging system and thus enables \emph{real-time} NLOS reconstruction.

\noindent\fangzhou{\textbf{Run time comparison}: Our model runs at \emph{11.8 FPS} on an NVIDIA V100 GPU. This far exceeds the data acquisition rate (5 FPS) of the imaging system and thus enables \emph{real-time} NLOS reconstruction. RSD and LFE run at 380 FPS and 30.8 FPS on the same GPU, whereas NeTF requires six hours of training on each scene.}

% \begin{table}[!t]
% \caption{Run time of different methods to reconstruct a single measurement.} \vspace{-0.5em}
% \label{runtime}
% \centering
% \begin{tabular}{c||ccccc}
% \hline
%      & RSD~\cite{liu2020phasor} & LFE~\cite{chen2020learned}  & NeTF~\cite{shen2021non}  & NeTF++ & Ours \\ \hline\hline
% Time & 3ms & 32ms & 6.3hr & 4.5hr  & 88ms \\ \hline
% \end{tabular}
% \end{table}

\subsection{Implementation Details}
For supervised training, We sample 4,096 rays uniformly at random from all target views. For unsupervised training, we sample one sensor location at a time and draw 4,096 rays originating from the sensor as discussed in Section~\ref{renderer}. We combine the two sets of ray samples in joint training. The models are trained for 50 epochs using the Adam optimizer~\cite{kingma2014adam} with a mini-batch size of 2 and a learning rate of 0.0001. Training on measurements of size $512 \times 128 \times 128$ takes three hours on a single NVIDIA V100 GPU and requires 8 GB of memory.

%% file: tex/06_exp_results.tex
\section{Experiments and Results}

We now present our experiments. We begin by introducing our datasets, evaluation metrics and imaging hardware. We then present our main results including intensity reconstruction, depth estimation and novel view synthesis. Finally, we probe the learned representations by training a classifier for object recognition, and end with an ablation study.

\subsection{Datasets and Metrics}\label{data}

\noindent \textbf{Synthetic Datasets}:
Our model and the learning based baseline of Chen et al.~\cite{chen2020learned} are exclusively trained on synthetic data. We simulate two large datasets of NLOS measurements and their respective multi-view target images using the transient rasterizer from~\cite{chen2020learned} with default parameters. 

The first \textbf{alphanumerics} dataset contains all lower and upper case letters from the English and Greek alphabets as well as digits 0 to 9. We randomly sample 25 poses for each of the 111 objects to synthesize a total of 2,775 samples. The transient measurements are $128\times128\times512$ in size. The target images are $256\times256$ and include 26 views as in Chen et al.~\cite{chen2020learned}. They both have a single brightness channel to match the real measurements. We create a training split and two test splits. The training set has 2,000 samples from 100 objects with 20 poses each. The ``Unseen Poses" test set includes the remaining five poses for each training object. The ``Unseen Objects" test set has the remaining 11 objects and their full set of poses. We report results on both test sets to evaluate the model's generalizability.

The second \textbf{motorbikes} dataset contains 6,925 samples rendered using 277 motorbikes from ShapeNet~\cite{chang2015shapenet}, where each motorbike is again rendered in 25 random poses. The transient measurements are $256\times256\times512$ in size with RGB color channels for fair comparison with Chen et al.~\cite{chen2020learned}. We generate training and test splits as before, with 5,000 samples from 250 motorbikes in the training set.

\fangzhou{We additionally simulate two small datasets for testing. The first \textbf{CMU} dataset includes six objects with complex geometry~\cite{tsai2019beyond}. The second has two objects from the \textbf{Z-NLOS} dataset~\cite{galindo19-NLOSDataset} and is previously used by NeTF. We use it in an ablation study to evaluate our transient rendering recipe.}

\smallskip
\noindent \textbf{Real Measurements}: We collect NLOS measurements of a few objects using our imaging hardware for model evaluation. The objects include digits ``2" and ``4" that are meant to evaluate the model's \emph{in-class} generalizability on real data, and ``chair", ``truck" and ``Teddy bear" that represent more challenging real-world objects unseen during training. The objects are placed approximately 1 meter away from the relay wall in various poses. Each measurement is paired with a reference image of the object for the qualitative assessment of reconstruction quality.

\smallskip
\noindent \textbf{Evaluation Metrics}: We compute root mean squared error (RMSE), peak signal-to-noise ratio (PSNR) and the structural similarity index measure (SSIM) to quantitatively evaluate the reconstruction results on synthetic data.

\begin{figure}[!t]
\centering
\includegraphics[width=0.9\linewidth]{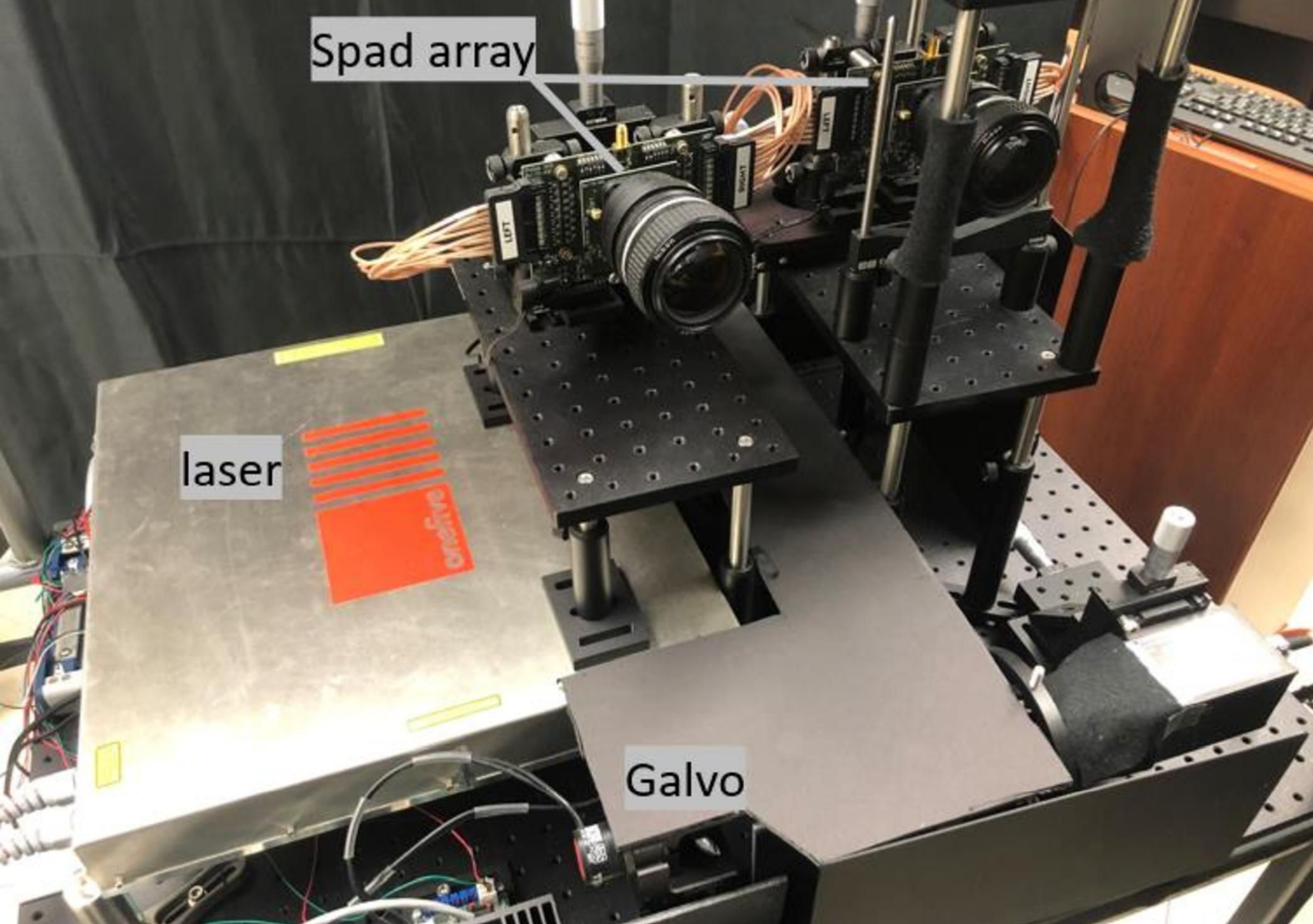} \vspace{-0.5em}
\caption{\textbf{Our imaging hardware}. Our prototype includes an ultrafast pulsed laser, two 1D SPAD arrays and a galvo for laser redirection.}
\vspace{-0.5em}
\label{hardware_figure}
\end{figure}

\subsection{Imaging Hardware}\label{hardware}

Our hardware prototype, depicted in Figure~\ref{hardware_figure}, consists of an ultrafast pulsed laser (OneFive Katana HP, 700~mW average power, 532~nm, 35~ps pulse width, 5~MHz repetition rate) and two 1D SPAD arrays each with 14 available pixels (75~ps FWHM)~\cite{Renna20}. The laser raster-scans a 190$\times$24 grid of locations that cover a 1.9m$\times$1.9m square on the wall. The SPAD arrays approximately co-localize and focus on the same 1cm$\times$9cm patch on the wall near the center of the scanned area. Photon counts from overlapping pixels are summed together, yielding 14 histograms per scanned location. The histograms have 768 bins with a temporal resolution of 32~ps. We apply the pixel remapping algorithm from Nam et al.~\cite{nam2021low} along with nearest-neighbor interpolation to convert the raw transients into a measurement of size $128\times128\times768$ for reconstruction.

% \subsection{Baseline Methods}\label{baseline}

% We extensively compare our approach with three strong baseline methods, namely RSD~\cite{liu2020phasor}, LFE~\cite{chen2020learned} and NeTF~\cite{shen2021non}.\fangzhou{We additionally compare with NeTF++, a variant of NeTF with our proposed volume rendering recipe (Section~\ref{renderer}).} RSD is the state-of-the-art physics based method for non-confocal NLOS reconstruction. Our encoder subsumes RSD as its feature propagation module. LFE, NeTF and NeTF++ are learning based and are related to our method in that they inject physics priors into the modeling. Operationally, RSD, LFE and our model are feed-forward methods that return a reconstruction in a single pass, whereas NeTF and NeTF++ are iterative optimization methods that require training a separate model for each measurement.

% We further qualitatively compare our method with three physics based methods developed for confocal reconstruction, namely filtered back-projection (FBP)~\cite{velten2012recovering}, light-cone transform (LCT)~\cite{o2018confocal} and f-k migration~\cite{lindell2019wave}. These methods can be adapted for non-confocal reconstruction at the expense of geometric distortion, reduced field of view and loss in image resolution. Past work has demonstrated that RSD is a much better alternative for physics based non-confocal reconstruction~\cite{liu2020phasor}.

\begin{figure*}[!t]
\centering
\includegraphics[width=\textwidth]{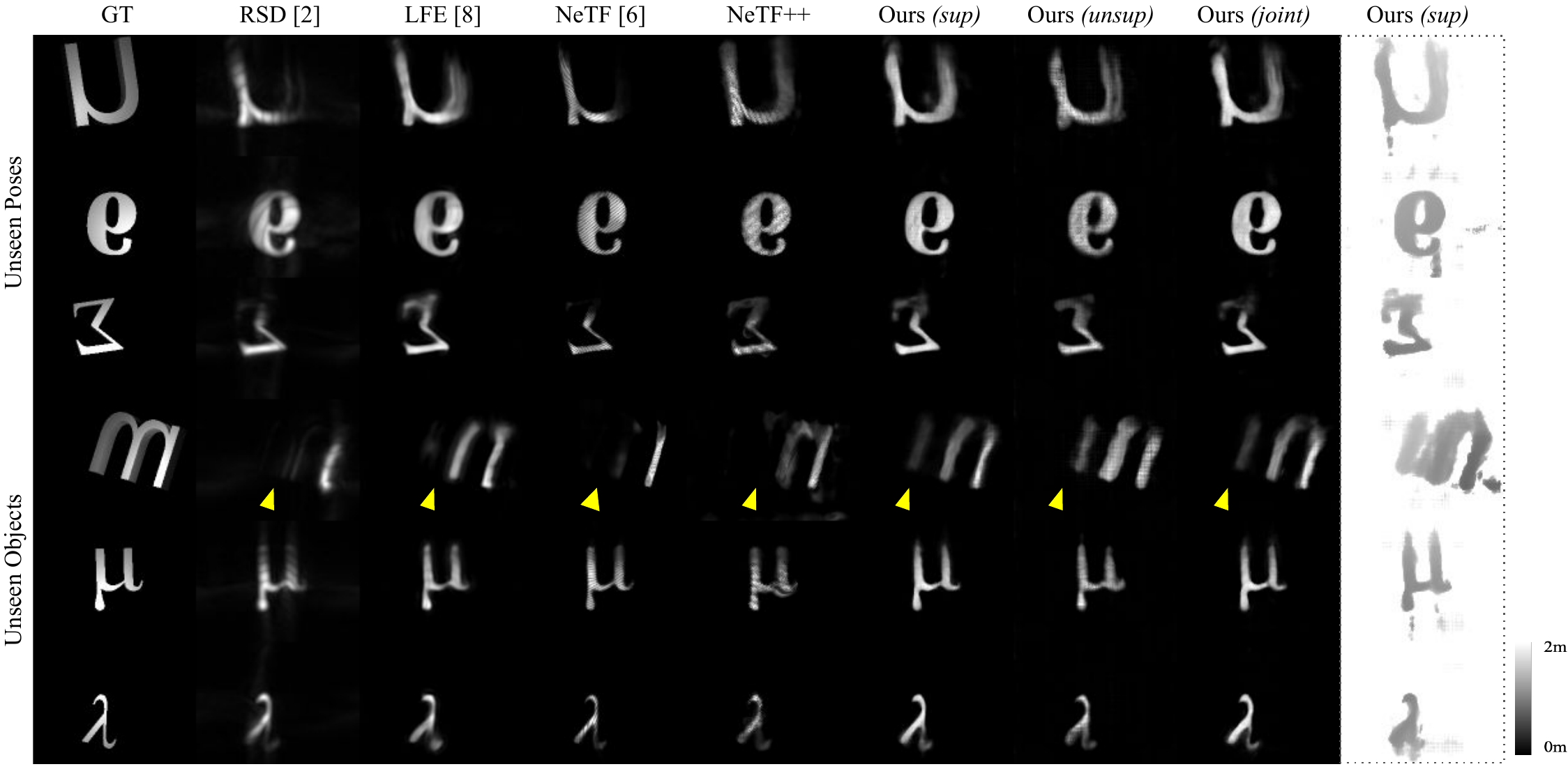} \vspace{-2em}
\caption{\fangzhou{\textbf{Reconstruction results on the \emph{alphanumerics} dataset.} Compared to the baselines, our method produces sharper reconstructions with finer details. Thanks to the learned scene priors, our method can infer scene content that are physically undetectable ({\color{Dandelion}{yellow}} arrows). The rightmost column presents estimated depth from our supervised model.}}
\vspace{-0.5em}
\label{alphanumerics_fig}
\end{figure*}

\begin{figure}[!t]
\centering
\includegraphics[width=\linewidth]{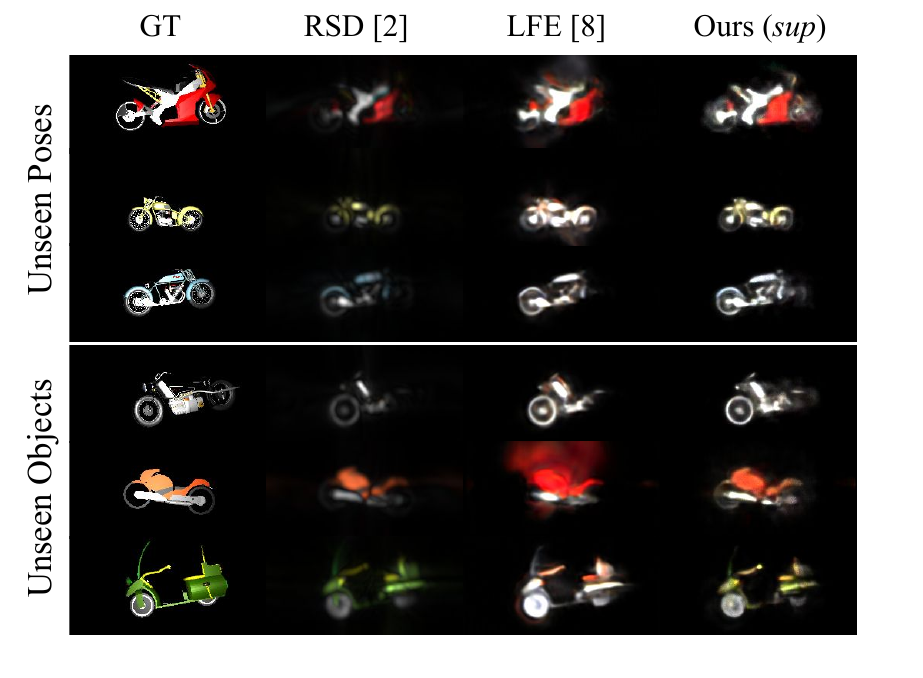} \vspace{-3em}
\caption{\textbf{Reconstruction results on the \emph{motorbikes} dataset.} Reconstructions from our method achieve better color balance and contain geometry details (e.g., wheels) missed by RSD and LFE.}
\vspace{-0.5em}
\label{motorbikes_fig}
\end{figure}

\begin{figure*}[!t]
\centering
\includegraphics[width=\linewidth]{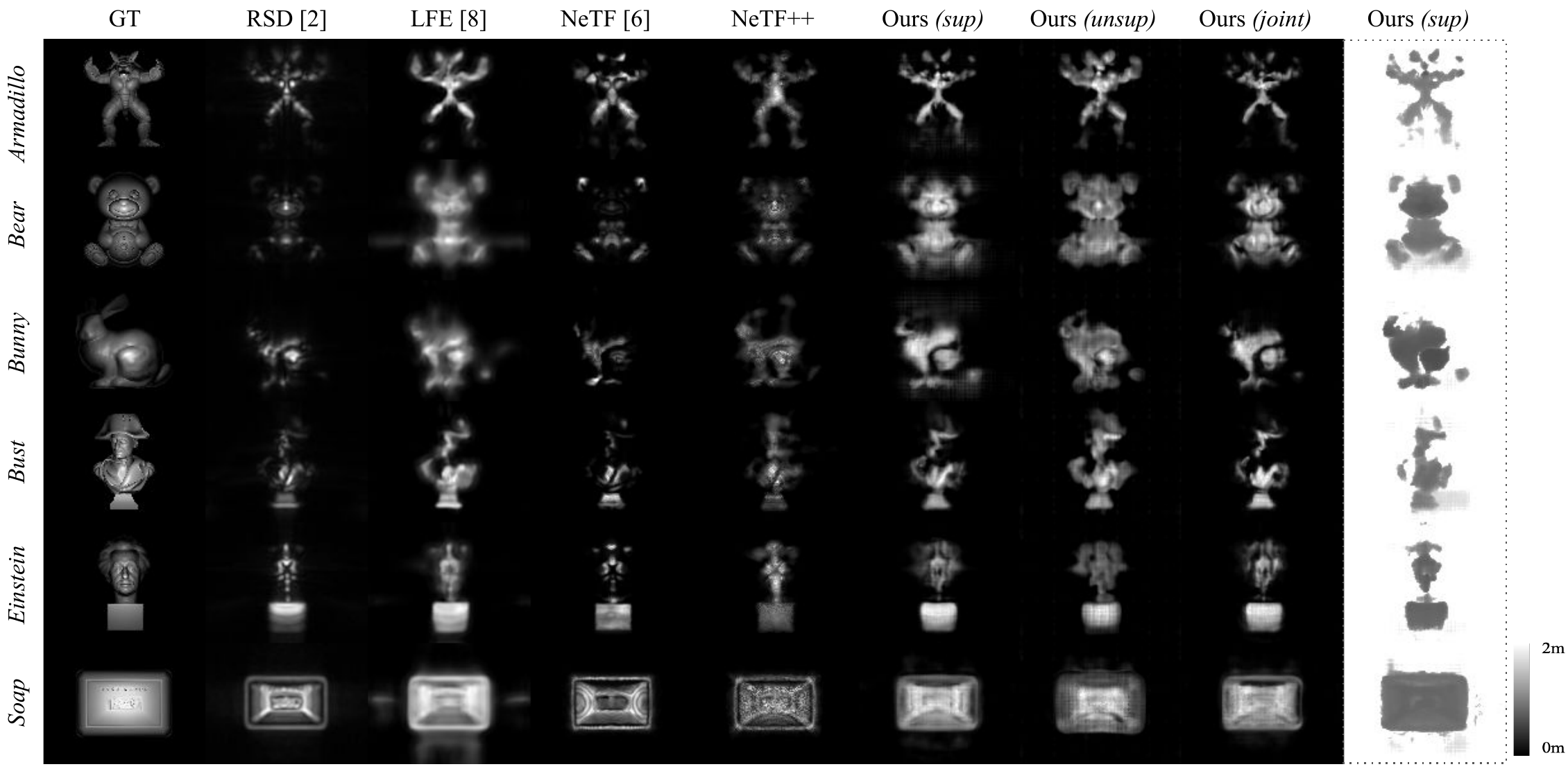} \vspace{-2em}
\caption{\fangzhou{\textbf{Reconstruction results on the \emph{CMU} dataset.} Despite trained on the \textbf{alphanumerics} dataset, our models generalize well on complex out-of-distribution shapes thanks to the strong regularization effect of the physics priors. The rightmost column presents estimated depth from our supervised model.}}
\vspace{-0.5em}
\label{cmu}
\end{figure*}

\begin{table}[!t]
\caption{Quantitative evaluation of feed-forward reconstruction methods on the \textit{full} \textbf{alphanumerics} test sets.} \vspace{-1em}
\label{alphanumerics_table}
\centering
\begin{tabular}{c||lll|lll}
\hline
\multirow{2}{*}{Methods} &
  \multicolumn{3}{|c|}{Unseen Poses} &
  \multicolumn{3}{c}{Unseen Objects} \\ 
  
  \cline{2-7} 
 &
  \multicolumn{1}{c}{RMSE} &
  \multicolumn{1}{c}{PSNR} &
  \multicolumn{1}{c|}{SSIM} &
  \multicolumn{1}{c}{RMSE} &
  \multicolumn{1}{c}{PSNR} &
  \multicolumn{1}{c}{SSIM} \\ \hline
  \hline
RSD~\cite{liu2020phasor}         & 0.095 & 20.83 & 0.395 & 0.094 & 20.92 & 0.407 \\
LFE~\cite{chen2020learned}         & 0.079 & 22.42 & 0.652 & 0.077 & 22.51 & 0.656 \\ \hline
Ours (\textit{sup})   & 0.062 & 24.60 & 0.816 & 0.060 & 24.80 & 0.821 \\
Ours (\textit{unsup}) & 0.093 & 20.97 & 0.709 & 0.092 & 21.02 & 0.726 \\
Ours (\textit{joint}) & \textbf{0.060} & \textbf{25.05} & \textbf{0.861} & \textbf{0.059} & \textbf{25.02} & \textbf{0.868} \\ \hline
\end{tabular}
\vspace{-0.5em}
\end{table}

\begin{table}[!t]
\caption{Quantitative evaluation of feed-forward reconstruction methods on the \textit{full} \textbf{motorbikes} test sets.} \vspace{-0.8em}
\label{motorbikes_table}
\centering
\begin{tabular}{cl||ccc|ccc}
\hline
\multicolumn{2}{c||}{\multirow{2}{*}{Method}} & \multicolumn{3}{c|}{Unseen Poses} & \multicolumn{3}{c}{Unseen Objects} \\ \cline{3-8} 
\multicolumn{2}{c||}{}          & RMSE  & PSNR   & SSIM   & RMSE  & PSNR   & SSIM  \\ \hline
\hline
\multicolumn{2}{c||}{RSD~\cite{liu2020phasor}}       & 0.087 & 21.87 & 0.632  & 0.079 & 22.65 & 0.648 \\
\multicolumn{2}{c||}{LFE~\cite{chen2020learned}}       & 0.092 & 21.09 & 0.854  & 0.088 & 21.41 & 0.854 \\ \hline
\multicolumn{2}{c||}{Ours (\textit{sup})} & \textbf{0.065} & \textbf{24.05} & \textbf{0.881} & \textbf{0.062} & \textbf{24.59} & \textbf{0.886} \\ \hline
\end{tabular}
\vspace{-0.5em}
\end{table}

\begin{table}[!t]
\caption{\fangzhou{Quantitative evaluation of all methods on \textit{selected}  \textbf{alphanumerics} (in-distribution) and \textbf{CMU} (out-of-distribution) test samples.}} \vspace{-1em}
\label{selected_table}
\centering
\begin{tabular}{c||ccc|ccc}
\hline
\multirow{2}{*}{Method} & \multicolumn{3}{c|}{Alphanumerics} & \multicolumn{3}{c}{CMU} \\ \cline{2-7} 
                        & RMSE      & PSNR       & SSIM      & RMSE   & PSNR   & SSIM  \\ \hline \hline
RSD~\cite{liu2020phasor}                     & 0.084     & 22.02     & 0.395     & 0.086  & 21.54 & 0.456 \\
LFE~\cite{chen2020learned}                     & 0.064     & 24.34     & 0.886     & 0.082  & 21.84 & 0.700 \\
NeTF~\cite{shen2021non}                    & 0.087     & 21.99     & 0.893     & 0.100  & 20.31 & 0.795 \\
NeTF++                  & 0.083     & 21.84     & 0.895     & 0.071  & \textbf{23.26} & \textbf{0.815} \\ \hline
Ours (\textit{sup})     & 0.059     & 24.94     & 0.905     & 0.076  & 22.46 & 0.799 \\
Ours (\textit{unsup})   & 0.073     & 23.12     & 0.833     & \textbf{0.070}  & 23.18 & 0.775 \\
Ours (\textit{joint})   & \textbf{0.057}     & \textbf{25.15}     & \textbf{0.896}     & 0.079  & 22.11 & 0.798 \\ \hline
\end{tabular}
\vspace{-0.5em}
\end{table}

\subsection{Comparison with RSD and LFE}\label{comp1}

Our method returns a reconstruction in a single forward pass. To this end, we first compare it to the feed-forward methods of RSD and LFE. RSD is the state-of-the-art physics based method for non-confocal NLOS reconstruction. Our encoder subsumes RSD as its feature propagation module. LFE is a learning based method trained using multi-view target images similar to our supervised model.
% We do not train LFE using ground-truth depth maps for fair comparison.

We train LFE and our model on \textbf{alphanumerics} and \textbf{motorbikes} and report quantitative results on their respective test sets (Table~\ref{alphanumerics_table} and~\ref{motorbikes_table}). Our supervised model consistently outperforms RSD and LFE by a wide margin. Our unsupervised model, which is solely trained to enforce cycle consistency, compares favorably against RSD. Importantly, our model achieves the best results with the joint objective, highlighting the benefit of our unified modeling approach. Qualitative results (Figure~\ref{alphanumerics_fig} and~\ref{motorbikes_fig}) demonstrate that our method reconstructs sharper contours and finer details, and in particular, is able to infer content that a physics based method cannot detect thanks to the learned scene priors.

\fangzhou{Moving forward, we investigate how models trained on \textbf{alphanumerics} generalize on the challenging out-of-distribution scenes from the \textbf{CMU} dataset. We present quantitative results averaged over the six available scenes in Table~\ref{selected_table} and qualitative results in Figure~\ref{cmu}. Our method outperforms RSD and LFE despite increasing complexity in scene geometry. Notably, our unsupervised model performs even better than the supervised LFE model thanks to the strong regularization effect of the physics priors.}

\begin{figure*}[!t]
\centering
\includegraphics[width=1.0\textwidth]{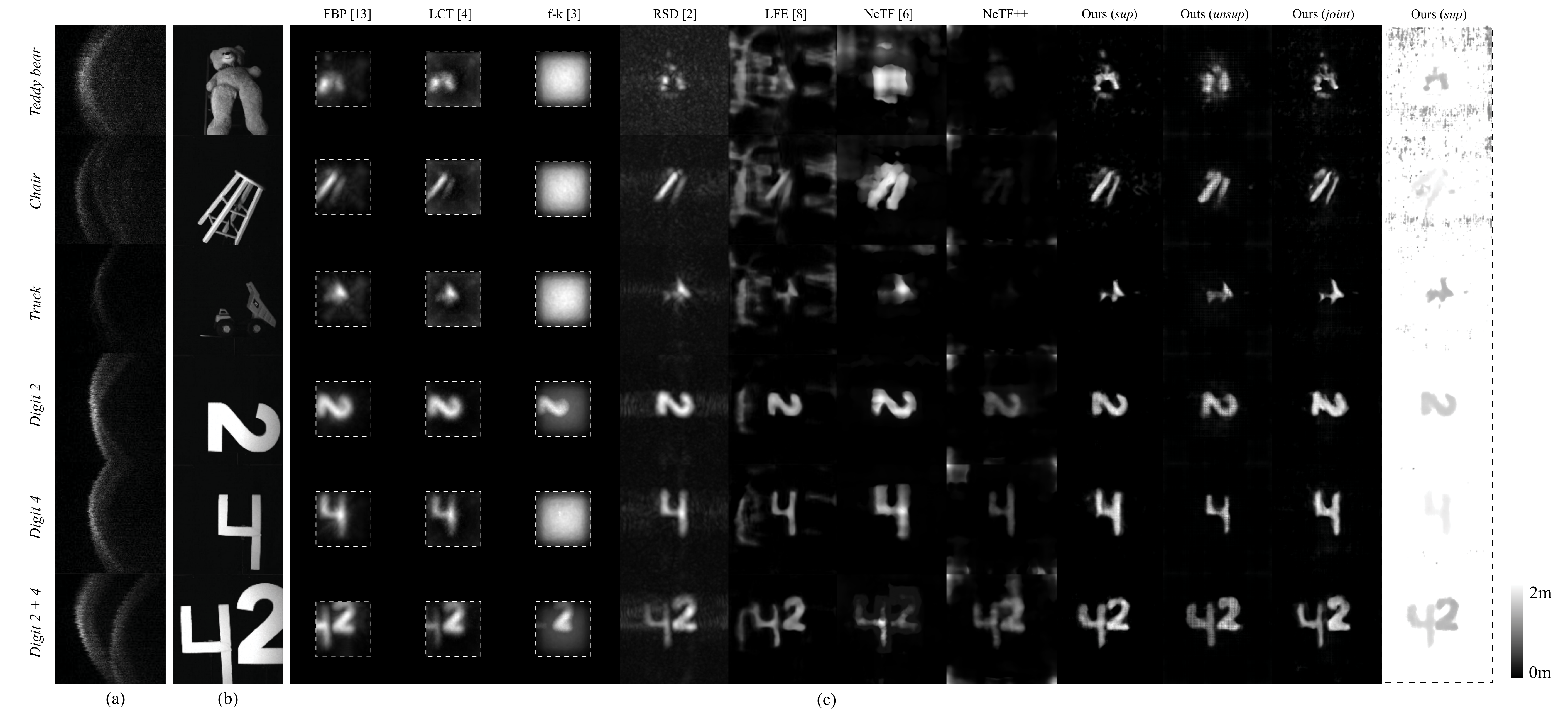}\vspace{-1em}
\caption{\textbf{Reconstruction results on real captures.} \textbf{(a)} An \textit{x-t} slice of the measurement volume. Note the roughness in the leading edge of returning photons due to pixel remapping, a key approximation that enables high-speed NLOS imaging. \textbf{(b)} A reference image of the hidden scene (not used for inference). \textbf{(c)}\fangzhou{Intensity and depth reconstruction. Our method is robust to approximations in light transport introduced by the high-speed imaging system and produces the best reconstructions on real captures. The rightmost column presents estimated depth from our supervised model.}}
% \caption{\textbf{Qualitative results on real measurements.} \textbf{(a)} An \textit{x-t} slice of the input measurement volume. Note the roughness in the leading edge of returning photons due to pixel remapping, a key approximation that enables high-speed NLOS imaging. \textbf{(b)} A reference image of the hidden scene (not used for inference). \textbf{(c)} Reconstruction results. FBP, LCT and f-k are confocal reconstruction methods that can be adapted for non-confocal reconstruction at the cost of lower-resolution output (dashed squares), reduced field of view and frequent failures. RSD is tailored for non-confocal reconstruction yet produces blurry and noisy results. Among the learning based methods, our approach demonstrates the strongest results on real measurements and has the capacity for depth estimation at inference time (rightmost column).}
\vspace{-0.5em}
\label{real_fig}
\end{figure*}

Finally and most importantly, we present qualitative comparisons on real captures in Figure~\ref{real_fig}. Our method equipped with the complementary physics priors produces strong reconstructions on scenes with varying degrees of complexity. In contrast, LFE with a ConvNet based neural renderer cannot accommodate the large domain gap between synthetic training data and real captures, resulting in distorted and noisy reconstructions even worse than RSD.

\subsection{Comparison with NeTF}\label{comp2}

Next, we compare our method with NeTF, a physics based iterative optimization method that requires training a separate model for each measurement. In addition, we introduce NeTF++, a variant of NeTF with our proposed rendering formulation (Section~\ref{renderer}).

\fangzhou{We train our model on \textbf{alphanumerics} and report results on test samples\footnote{It is infeasible to train NeTF and NeTF++ on all test samples. We hence report results averaged over six random test scenes.} of the same dataset as well as the \textbf{CMU} dataset. Our method outperforms NeTF and NeTF++ on \textbf{alphanumerics} (Table~\ref{selected_table} and Figure~\ref{alphanumerics_fig}) while supporting fast feed-forward inference. We attribute this to the learned scene priors that allow our model to reason beyond individual measurements. However, the scene priors inevitably fall short in the presence of a domain shift, which explains why NeTF++ outperforms our method on the \textbf{CMU} dataset (Table~\ref{selected_table} and Figure~\ref{cmu}). Meanwhile, NeTF++ and our method that share the same transient renderer compare favorably against NeTF. This evidently shows the strength of our proposed rendering recipe.} %which we further investigate in our ablation study.

On real captures, NeTF and NeTF++ are sensitive to the approximations in light transport not captured by their underlying image formation model. As a result, their reconstructions are prone to artifacts and sometimes incomplete especially for complex scenes. On the other hand, our method reliably reconstructs simple objects and behaves strongly on complex scenes. 

% \subsection{Unsupervised Learning Results}\label{unsup}

% In the unsupervised setting, we compare our method to RSD and report metrics averaged over the test sets of \textbf{alphanumerics} (Table~\ref{alphanumerics_table}). Further, we compare our method to NeTF and report metrics on individual scenes from the same test sets (Figure~\ref{unsup_netf}), noting that it is infeasible to train NeTF on every single measurement. Finally, we report qualitative results on synthetic and real measurements in Figure~\ref{ch1_128_sy} and~\ref{real_fig}.

% Our model, which is solely trained to enforce cycle consistency, consistently outperforms RSD by a clear margin. Our method also compares favorably against NeTF on individual scenes, even though NeTF spends \emph{days} to optimize a \emph{single} reconstruction. NeTF is also sensitive to approximations in the imaging model and produces even worse reconstructions on real data with erroneous geometry at borders. Our method, on the other hand, yields the best reconstructions among all unsupervised competitors thanks to its built-in robustness.

\subsection{Comparison with FBP, LC and f-k}\label{comp3}

Finally, we qualitatively compare our method with three physics based methods developed for confocal reconstruction, namely filtered back-projection (FBP)~\cite{velten2012recovering}, light-cone transform (LCT)~\cite{o2018confocal} and f-k migration~\cite{lindell2019wave}. These methods can be adapted for non-confocal reconstruction at the expense of geometric distortion, reduced field of view and loss in image resolution. Past work has demonstrated that RSD is a much better alternative for physics based non-confocal reconstruction~\cite{liu2020phasor} and our results on real captures (Figure~\ref{real_fig}) point to the same conclusion.

\subsection{Depth Estimation}\label{depth}

We have so far presented results on the reconstruction of object appearance. We now demonstrate that our method, \emph{depite not being trained for depth estimation}, is able to predict plausible depth thanks to our versatile scene representation and volume rendering framework. In contrast, it is non-trivial to engineer RSD or an LFE model trained solely on 2D images for depth estimation. We report depth estimation results from our supervised model in the rightmost columns of Figure~\ref{alphanumerics_fig},~\ref{cmu} and~\ref{real_fig}. Our method works equally well on synthetic and real data, producing plausible depth prediction for front-facing scenes and uncovers depth of the visible surfaces for objects with complex geometry.

% Specifically, we compute the expected distance where a ray collides with a scene point
% \begin{equation}
%     \hat{d}(\mathbf{r}) = \int_0^\infty T(u)\sigma(\mathbf{r}(u))u\mathrm{d}u.
% \end{equation}

% \begin{figure}[h]
% \centering
% \includegraphics[width=\linewidth]{figures/depth.pdf}\vspace{-0.5em}
% \caption{\textbf{Depth estimation results on the \textit{alphanumerics} dataset.} Our method produces sharp and accurate depths despite not being trained for depth prediction.}
% \label{sup_depth}
% \end{figure}

\begin{figure}[!t]
\centering
\includegraphics[width=1\linewidth]{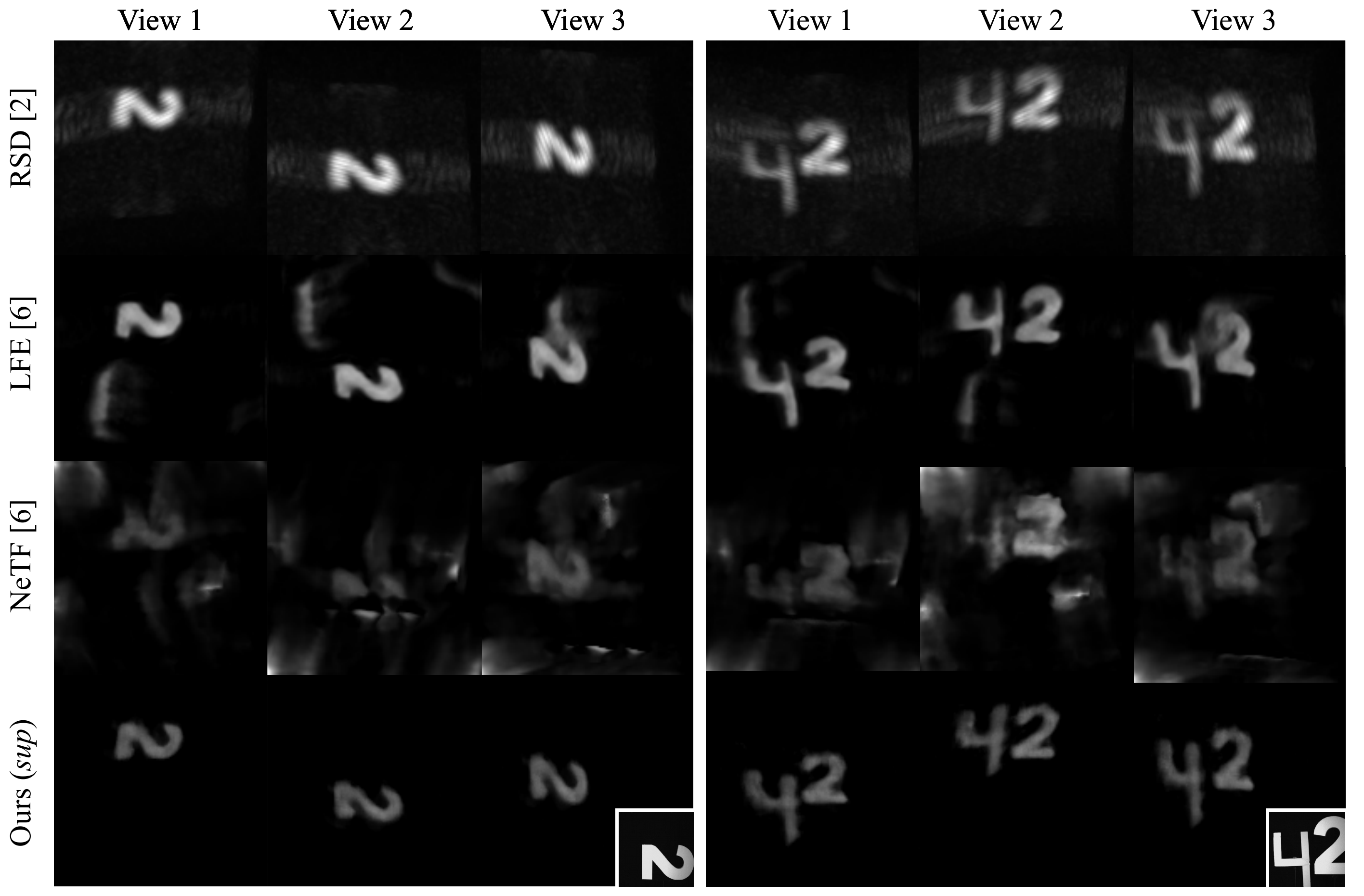} \vspace{-2em}
\caption{\textbf{Novel view synthesis.} \fangzhou{Our method can render intensity images beyond the frontal view. Novel view synthesis shows that our reconstructions capture accurate 3D geometry.} A reference view of the hidden scene is displayed in the inset.} \vspace{-0.5em}
\label{nvs_img}
\end{figure}

\subsection{Novel View Synthesis}\label{nvs}

NLOS reconstruction algorithms are often evaluated on the  \emph{frontal} view of the scene observed from behind the wall. We hypothesize that accurate 3D reconstruction of a hidden object would facilitate view synthesis beyond the frontal view. We thus task RSD, LFE, NeTF and our method for novel view synthesis to assess the quality of 3D reconstruction.

We render three random views given real measurements of two digits (``2'' and ``4''), and demonstrate the results of different methods in Figure~\ref{nvs_img}. Our method produces crispy, artifact-free renderings compared to the noisy and blurry outputs of the baseline methods.

\begin{table}[!t]
\caption{Classification results on \textbf{alphanumerics} using learned features.} \vspace{-1em}
\label{feature_table}
\centering
\begin{tabular}{ccccccc}
\cline{1-6}
%\hline
\multicolumn{1}{c||}{}   & \multirow{2}{*}{RSD~\cite{liu2020phasor}} & \multirow{2}{*}{LFE~\cite{chen2020learned}} & Ours           & Ours             & Ours             &  \\
\multicolumn{1}{c||}{}   &                      &                      & (\textit{sup}) & (\textit{unsup}) & (\textit{joint}) &  \\ \hline \hline
\multicolumn{1}{c||}{Accuracy} &82.2\% & 81.0\% & \textbf{85.4\%} & \textbf{85.8\%}  & \fangzhou{\textbf{85.2\%}}               &  \\ \cline{1-6}
                        &                      &                      &                &                  &                  & 
\end{tabular}
\vspace{-1em}
\end{table}

% \begin{table}
% \caption{Classification results on \textbf{alphanumerics} using learned features.} \vspace{-0.5em}
% \label{feature_table}
% \centering
% \begin{tabular}{l||r} 
% \hline
% Feature              & \multicolumn{1}{l}{Accuracy}  \\ 
% \hline \hline
% RSD                  & 82.2\%                        \\
% LFE                  & 81.0\%                        \\
% \hline
% Ours (\textit{sup})   & \textbf{85.4\%}               \\
% Ours (\textit{unsup}) & \textbf{85.8\%}               \\
% Ours (\textit{joint}) & \textbf{85.2\%}               \\
% \hline
% \end{tabular}
% \end{table}

\subsection{Object Classification}\label{feat}

We further compare our method to RSD and LFE by adapting the learned features for object classification. This surrogate task helps reveal the model's capacity to encode scene priors useful for downstream recognition tasks. 

Recall that both our method and LFE encode and propagate a transient measurement to a spatial feature cube. For both models trained on the \textbf{alphanumerics} dataset, we project the feature cube to a 2D feature map by taking the per-channel maximum along the depth axis. We similarly project the volumetric reconstruction of RSD and interpret the 2D projection as a feature map. We then train a ResNet-18~\cite{he2016deep} on the 2D features for 100-way alphanumeric classification. All classifiers are trained for 50 epochs with a mini-batch size of 32 using stochastic gradient descent (SGD) with a learning rate of 0.1, a momentum of 0.9 and a weight decay of $5\times10^{-4}$. We report classification accuracy on the ``Unseen Poses" test set (Table~\ref{feature_table}), along with predicted confidences for real measurements of digits ``2" and ``4" (Figure~\ref{classification_real}).

Classifiers using features from our models achieve $85.4\%$ (supervised) and $85.8\%$ (unsupervised) test accuracy, outperforming those using RSD and LFE features by more than $3\%$ and $4\%$, respectively. Moreover, our classifier predicts the correct labels with high confidence for the real measurements of ``2", whereas the baselines confuse ``2" with the morphologically similar letters ``S" and ``Z" and digit ``3". All classifiers fail on the measurement of ``4" since it looks quite different from the digit ``4" in the training set (see the inset), yet classifiers using our features make the most plausible prediction. Our results suggest that our encoder learns scene priors with more discriminative power.

\subsection{Ablation study}
\noindent \textbf{Total Variation Prior}: Our training objectives include a total variation term to encourage sparsity in scene opacity (Equation~\ref{tv_eq}). A model trained without the regularizer renders decent intensity images but introduces spurious density in empty space (Figure~\ref{tv_ablation}).

\smallskip
\noindent \textbf{Transient Rendering Recipe}:
NeTF models reflectance at every scene point. Unfortunately, the transmittance term in their rendering equation has to be omitted for tractability of numerical integration. Our renderer instead models outgoing radiance without factoring it into illumination and reflectance. Importantly, this enables transmittance estimation with a tractable sampling strategy (Figure~\ref{sampling}). To disentangle the effect of our implementation decisions, we omit transmittance in NeRF++ and call this variant NeRF+, from which we recover NeTF by modeling reflectance rather than radiance. We compare NeTF, NeTF+ and NeTF++ on two scenes from the \textbf{Z-NLOS} dataset~\cite{galindo19-NLOSDataset}. Results in Figure~\ref{netf_ablation} show that proper evaluation of transmittance is key to high-quality reconstruction, while the choice of modeling radiance as opposed to reflectance itself is less important.

% \begin{table}[!t]
% \renewcommand{\arraystretch}{1.3}
% \caption{An Example of a Table}
% \centering
% \begin{tabular}{c||c|c|c}
% \hline
% One & Two & One & Two\\
% \hline\hline
% Three & Four &Three & Four\\
% \hline
% Three & Four &Three & Four\\
% \hline
% Three & Four &Three & Four\\
% \hline
% \end{tabular}
% \end{table}